\documentclass[a4paper,12pt]{article}
\usepackage{graphicx,amssymb,amstext,amsmath,txfonts}
\usepackage{xcolor, yfonts, tcolorbox}
\usepackage{floatflt}
\usepackage{comment}
\usepackage{footnote}
\usepackage{float}
\usepackage{esint}
\usepackage{subcaption}
\usepackage{mathtools}
\usepackage{commath}
\usepackage{array}
\usepackage{boldline}
\usepackage{multirow}
\usepackage{braket}
\usepackage{arydshln}
\usepackage{bigdelim}
\usepackage{hyperref}
\usepackage{simpler-wick}
\usepackage[style=phys,hyperref=true,articletitle=true,biblabel=brackets,chaptertitle=false]{biblatex}
\definecolor{cbl}{rgb}{0,0,1}                

\newcommand{\bc}{\begin{center}}
\newcommand{\ec}{\end{center}}
\def\ba#1{\begin{array}{#1}\displaystyle}
\newcommand{\ea}{\end{array}}

\newcommand{\beq}{\begin{equation}}
\newcommand{\eeq}{\end{equation}}
\newcommand{\beqa}{\begin{eqnarray}}
\newcommand{\eeqa}{\end{eqnarray}}

\newcommand{\bi}{\begin{itemize}}
\newcommand{\ei}{\end{itemize}}
\newcommand{\TTb}{\mathrm{T\bar{T}}}

\newcommand{\varep}{\varepsilon}

\newcommand{\Tr}{{\rm Tr}}

\def \be {\begin{equation}} 
\def \ee {\end{equation}} 
\def \l {\left(} 
\def \r {\right)}

\newcommand{\MM}[1]{\textcolor{red}{#1}}

\begin{document}
\begin{titlepage}
\vspace{0.2cm}
\begin{center}

{\large{\bf{Expectation Values of Conserved Charges in Integrable Quantum Field Theories out of Thermal Equilibrium}}}

\vspace{0.8cm} 
{\large Michele Mazzoni{\Large $^{{\color{blue}\spadesuit}}$},  Riccardo Travaglino{\Large $^{{\color{teal}\star}}$} and Olalla A. Castro-Alvaredo{\Large $^{\LARGE\color{purple}\varheartsuit}$}}

\vspace{0.8cm}
{
{\Large $^{{{\color{blue}\spadesuit}}{{\color{purple}\varheartsuit}}}$} Department of Mathematics, City, University of London, \\ 10 Northampton Square EC1V 0HB, UK\\
\medskip
{\Large $^{\color{teal} \star}$}  SISSA, via Bonomea 265, 34136 Trieste, Italy
}
\end{center}

\medskip
\medskip
\medskip
\medskip
\noindent
In this work we present a computation of the averages of conserved charge densities and currents of (1+1)-dimensional Integrable Quantum Field Theories in Generalised Gibbs Ensembles. Our approach is based on the quasi-particle description provided by the Thermodynamic Bethe Ansatz combined with the principles of Generalised Hydrodynamics, and we focus on Non-Equilibrium Steady State averages. When considering the ultraviolet (i.e. high temperature) limit of such averages, we recover the famous result by Bernard and Doyon (2012) for the energy current and density in Conformal Field Theories, and we extend it to conserved quantities with spin $s > 1$. We show that their averages are proportional to $T_L^{s+1}\pm T_R^{s+1}$, with $T_L$, $T_R$ the temperatures of two asymptotic thermal reservoirs. The same power law is obtained when considering some non-thermal generalised Gibbs states. In Conformal Field Theory, the power law is a consequence of the transformation properties of conserved charge operators, while the proportionality coefficient depends on the spin of the operator and on the central charge of the theory. We present an exact analytic expression for this coefficient in the case of a massive free fermion. At equilibrium, proportionality of spin-$s$ density averages to $T^{s+1}$ can be thought of as a generalisation of Stefan-Boltzmann's law, which states that the energy per unit surface area radiated by a black body scales as $T^4$. 
\noindent 
\medskip
\medskip
\medskip
\medskip


\vfill
\noindent 
{\Large $^{{{\color{blue}\spadesuit}}}$} michele.mazzoni.2@city.ac.uk\\
{\Large $^{\color{teal} \star}$} rtravagl@sissa.it\\
{\Large $^{{{\color{purple}\varheartsuit}}}$} o.castro-alvaredo@city.ac.uk

\hfill \today

\end{titlepage}
\tableofcontents

\section{Introduction}
The Physics of many-body systems out of thermodynamic equilibrium has drawn an increasing amount of attention in recent years. The experimental finding that systems with an extensive number of conserved quantities do not display traditional thermalisation properties \cite{kinoshita2006quantum} has brought the concept of a Generalised Gibbs Ensemble (GGE) to prominence \cite{GGE_definition}. A GGE is a statistical ensemble that can describe the large-time average properties of (1+1)-dimensional integrable systems. The description of the non-equilibrium dynamics at mesoscopic scale in the presence of infinitely many conserved quantities (and, therefore, infinitely many currents), in which the Gibbs states are replaced by GGEs, is captured by the theory of Generalised Hydrodynamics (GHD) \cite{castro2016emergent,bertini_ghd}. The assumption at the basis of GHD is that of local entropy maximisation. This is the idea that entropy maximisation occurs within  fluid cells which contain a macroscopic number of degrees of freedom but are still small enough so that the variation of all observables with respect to the coordinates $(x,t)$ is smooth when moving between neighbouring cells. This is the scale at which the GHD description is effective (see also the reviews \cite{doyon2020lecture,ESSLER_ghd_review}).

The local entropy maximisation principle makes it possible to move the $(x,t)$-dependence from a local observables $\mathcal{O}$ to the Lagrange multipliers $\boldsymbol{\beta}=(\beta_1, \beta_2,\dots)$ that describe the state: 
\be 
\langle \mathcal{O}(x,t) \rangle \approx \langle \mathcal{O}(0,0) \rangle_{\boldsymbol{\beta}(x,t)} = \Tr[\mathcal{O}(0,0)\rho_{GGE}],
\ee
where the GGE state is
\be 
\label{GGE_rho_general}
\rho_{GGE} = \frac{e^{-\sum_s \beta_s(x,t) Q_s}}{\Tr\left[e^{-\sum_s \beta_s(x,t) Q_s}\right]}.
\ee
The quantities $Q_s$ are the conserved charges of the model, whose associated densities $q_s(x,t)$ satisfy the continuity equations
\be
\label{continuity_equation}
\partial_t q_s(x,t) + \partial_x j_s(x,t) = 0, \quad Q_s = \int \mathrm{d}x \, q_s (x,t)\,,
\ee
with $j_s(x,t)$ the corresponding currents. 
The charges are labelled by the value of the spin $s$, which is integer for all the local conserved quantities that can be constructed in an Integrable Quantum Field Theory (IQFT). However, the full description of a GGE requires the inclusion of quasi-local charges, which are typically associated with fractional spins \cite{failure, ProsenXXZ, ilievski2016quasilocal, vernier2017quasilocal}. 

In the Euler approximation currents are ballistic and by integrating \eqref{continuity_equation} over a fluid cell, one obtains Euler's equations for the averages:
\be
\label{Euler_equation_fluid_cell}
\quad \partial_t \langle q_s(x,t) \rangle + \partial_x \langle j_s(x,t) \rangle= 0.
\ee
One of the simplest yet most predictive situations in which the GHD equations can be solved is the partitioning protocol, the same considered in the original papers \cite{castro2016emergent, bertini_ghd}. In this setting, the system is in a homogeneous state almost everywhere: at time $t=0$, two thermal reservoirs at temperatures $T_R$ and $T_L$ respectively on the right and left semi-infinite half-lines are joined at $x=0$ and the system is then let evolve. As integrability forbids thermalisation, one observes ballistic currents and hydrodynamic correlation spreading within a light-cone centered at $(x,t)=(0,0)$. The energy current and density in the Non-Equilibrium Steady State (NESS) which is formed at $x=0$ for large times were computed in \cite{bernard2012energy, Bernard_2014, bernard2016conformal} for Conformal Field Theory (CFT). As the mean energy density in a CFT at finite temperature $T$ is $\mathbf{h}  = \frac{\pi c T^2}{6}$, where $c$ is the central charge of the theory\footnote{If the CFT is not unitary, the central charge is replaced by the effective central charge $c_{\rm eff}\equiv c-24\Delta$, where $\Delta$ is the smallest conformal dimension of a primary field of the theory (which could take a negative value).}, the energy current and charge average in the NESS are
\be
\label{CFT_energy_results}
\langle j_e \rangle = \frac{\pi c}{12}\l T_L^2-T_R^2\r, \quad \langle q_e\rangle = \frac{\pi c}{12}\l T_L^2+T_R^2\r. 
\ee
These formulae were obtained using results from finite size CFT at thermal equilibrium and have been thereafter verified both numerically and analytically for many situations \cite{NatureBe,KM,Luca,Luca2,LucaBe,Marton, Perfect,Lorenzo1,Lorenzo2}. The aim of this paper is to study the expectation values of the densities and related currents of higher-spin charges in IQFTs in the NESS formed after a partitioning protocol. We will use for this the Thermodynamic Bethe Ansatz (TBA) approach and its generalisation to GGEs \cite{Fioretto:2009yq,Mossel2012generalized}. We show that in the conformal limit the averages of local conserved charges and currents of generic spin satisfy:
\begin{align}
\begin{split}
\label{higher_spin_temperature_dependence}
 &\langle q_s \rangle \simeq \langle j_{-s} \rangle \propto \l T_L^{s+1} + T_R^{s+1}\r, \\
 &\langle j_s \rangle \simeq \langle q_{-s} \rangle \propto  \l T_L^{s+1} - T_R^{s+1}  \r,
\end{split}
\end{align}
where $s$ is the spin and the signs \lq\lq + \rq\rq \, and \lq\lq -\rq\rq \, refer to the parity (even/odd) of the corresponding charge eigenvalues w.r.t. the rapidity variable. The proportionality factors are in general theory-dependent and non-trivial. From our derivation we recover the correct coefficient $\frac{\pi c}{12}$ in \eqref{CFT_energy_results}, which corresponds to the case $s=1$ in the notation above. 
For higher spins, the generalisations of this coefficient can be regarded as higher-spin versions of the scaling function of the finite-temperature QFT \cite{Zamolodchikov1991tba, klassen1990, Klassen1991}, a dimensionless function obtained from the free energy density. Moreover, The TBA formalism for interacting theories allows us to extend this result to the case in which the spectrum of the theory contains several distinct stable particles.

The main advantage of the TBA, however, is that it provides a way to study the partitioning protocol with different asymptotic boundary conditions, namely when the reservoirs are characterised by GGEs rather than by thermal (Gibbs) states. In particular, we obtain exact results (in the conformal limit) when the state of the left and right reservoir are of the form
\be
\label{spin_s_state_rho}
\rho_{L/R} \sim e^{-(T^{-1}_{L/R})^s Q_s},
\ee
for some conserved charge of spin $s$. With this choice of the potential coupled to $Q_s$, the dependence on the temperatures $T_{L/R}$ is again given by \eqref{higher_spin_temperature_dependence} and the coefficients provide yet another generalisation of the QFT scaling function. 

Although the proportionality coefficients in \eqref{higher_spin_temperature_dependence}, as well as those arising with the choice \eqref{spin_s_state_rho}, are in general difficult to obtain analytically, there  is one specific situation in which they are exactly proportional to the standard scaling function -- that is, they flow to the effective central charge in the conformal limit. This happens when the time evolution of the system is ruled by a conserved charge of spin $s$ and we compute the expectation value of the density/current of the same charge. In doing so, the usual definition of the averages is revised to accommodate the notion of generalised time, associated to a \lq\lq generalised Hamiltonian\rq\rq $Q_s$. This is consistent with the CFT result \eqref{CFT_energy_results}, which correspond to the case of spin $s=1$ and where the generator of time evolution is the Hamiltonian.
\medskip 

This paper is organised as follows. In Section \ref{Section: preliminaries} we review the TBA and GHD results that we  will use throughout the rest of the work. We focus on systems  at equilibrium and on the partitioning protocol. In Section \ref{Section: main results} we present our main results, that is, the expressions of NESS averages of conserved charge densities and currents in the conformal limit, considering different asymptotic conditions. The free fermion case is discussed in Section \ref{Exact results for the massive free fermion}, where we provide exact expressions for all averages for arbitrary mass and temperature. In Section \ref{Numerical results} we test our conformal point predictions numerically for the sinh-Gordon and the Lee-Yang IQFTs.  We conclude in Section \ref{Conclusions and Outlook}. In Appendix \ref{appendix:Useful bounds on the effective velocity} we prove some bounds which are useful to obtain our main results for the partitioning protocol. Appendix \ref{appendix:Spin-dependent scaling function} contains a derivation of a class of spin-dependent scaling functions and Appendix \ref{Appendix:Many-particle theories} contains the generalisation of our results to IQFTs with a simple many-particle spectrum. We leave some details of the free-fermion calculations to Appendix \ref{Appendix:Free fermion integrals and finite temperature corrections}. 

\section{IQFT in a GGE and the Partitioning Protocol}
\label{Section: preliminaries}
In this Section we present the TBA equations that constitute the starting point for our derivation of the current averages. The TBA equations for relativistic IQFTs in thermal states were famously first derived in \cite{Zamolodchikov1991tba}, while the generalisation to homogeneous GGEs, that is states of the form \eqref{GGE_rho_general} with constant potentials $\beta_s$, was introduced in \cite{Fioretto:2009yq,Mossel2012generalized}. The equations we present here are those describing both a thermal state and states of the form \eqref{spin_s_state_rho}, in which there is a single non-vanishing potential which is a power of the inverse temperature. We call the latter a spin-$s$ state. Next, we show how the TBA quantities are used to describe expectation values of charge densities and current densities in GHD \cite{castro2016emergent}. We also present a useful original relation between the eigenvalues of a higher-spin conserved charge and the derivative of the TBA pseudoenergy in a spin-$s$ state. In the final part of this Section we present the solution of the partitioning protocol.

\subsection{TBA in Homogeneous GGEs}
The scattering theory of IQFTs in 1+1 space-time dimensions is described by picking a basis of asymptotic states labelled by the rapidities of the particles $\vartheta_i$ and their quantum numbers $a_i$, either in the remote past or in the remote future:
\be
\ket{\vartheta_1,\vartheta_2,\dots,\vartheta_n}^{\rm in/out}_{a_1,a_2,\dots,a_n}, \qquad 
\begin{cases}
    &\vartheta_1 < \dots < \vartheta_n, \quad \text{out state} \\
     &\vartheta_1 > \dots > \vartheta_n, \quad \text{in state}
\end{cases}.  
\ee
These asymptotic states are eigenstates of the charges defined in \eqref{continuity_equation}. If the theory is parity-invariant, it is possible to combine these charges in such a way that their eigenvalues have well-defined parity under a rapidity inversion. The action of the even and odd spin-$s$ charges, denoted respectively by $Q_s$ and $Q_{-s}$, $s > 0$, is given by
\be
Q_{\pm s}\ket{\vartheta_1,\dots,\vartheta_n}_{a_1,\dots,a_n} = \sum_{i=1}^n h_{a_i,\pm s} (\vartheta_i)\ket{\vartheta_1,\dots,\vartheta_n}_{a_1,\dots,a_n},
\ee
with:
\be
\label{eigenvalues_even_odd_definition}
h_{a_i,s} (\vartheta) = \chi_{a_i}^s m_{a_i}^s \cosh(s\vartheta), \quad h_{a_i,-s} (\vartheta) = \chi_{a_i}^s m_{a_i}^s \sinh(s\vartheta)\,.
\ee 
The quantities $m_{a_i}$ are the masses of the particles in the theory and the numbers $\chi_{a_i}$ are determined using a bootstrap approach \cite{Zamolodchikov:1989hfa}. In the case of a single particle type $a$, $\chi_a=1$. This is the situation we consider in this paper, leaving the discussion of many-particle theories to Appendix \ref{Appendix:Many-particle theories}. The $s=1$ charges are the Hamiltonian and the momentum, $Q_1=H$, $Q_{-1}=P$. For a particle with mass $m$ their eigenvalues are
\be
\label{eigenvalues_energy_momentum}
e(\vartheta) \equiv h_1(\vartheta) = m \cosh (\vartheta), \quad p(\vartheta) \equiv h_{-1}(\vartheta) = m \sinh (\vartheta)\,.
\ee
The TBA equations determine the thermodynamics of an IQFT at equilibrium at a finite temperature $T$. Let us assume that the spectrum of the IQFT consists of a single fermionic particle of mass $m$ and that the self-interaction is ruled by a scattering matrix $S(\vartheta)$. In this case there is a single TBA equation, a nonlinear integral equation for the pseudoenergy $\varepsilon (\vartheta)$:
 \be
 \label{TBA_eq_thermal}
 \varepsilon(\vartheta)= \beta m \cosh(\vartheta) - \varphi * L (\vartheta),
 \ee
where $\beta=T^{-1}$ is the inverse temperature,
\be
\varphi(\vartheta) = -i \frac{\partial}{\partial\vartheta}\log S(\vartheta),
\ee
is the scattering kernel of the theory, the function $L(\vartheta)$ is given by
\be
L(\vartheta) = \log \l 1 + e^{-\varepsilon(\vartheta)} \r,
\ee
and $*$ indicates convolution which is defined with a prefactor $(2\pi)^{-1}$. The TBA equation follows from the functional minimisation of the free energy subject to a constraint relating the density of available states $\rho_s(\vartheta)$ and the density of occupied states $\rho_p(\vartheta)$ at rapidity $\vartheta \in \mathbb{R}$. The occupation function $n(\vartheta) \equiv \frac{\rho_p(\vartheta)}{\rho_s(\vartheta)}$ is related to the pseudoenergy by
 \be
 \label{filling_function_thermal_state}
 n(\vartheta) = \frac{1}{1+e^{\varepsilon(\vartheta)}}.
 \ee
Once the TBA equation is solved, one can compute the finite-temperature ground state energy of the system as:
\be
E(\beta) = -\frac{1}{2\pi} \int\frac{\mathrm{d}\vartheta}{2\pi} e(\vartheta) L(\vartheta),
\ee
and the dimensionless scaling function is given by
\be
c(r) = -\frac{6\beta E (\beta)}{\pi} = \frac{3 r}{\pi^2} \int \mathrm{d}\vartheta \cosh (\vartheta) L(\vartheta).
\ee
The scaling function depends solely on the dimensionless variable $r=m\beta$, the value of which determines the \lq\lq position\rq\rq of the theory along the renormalisation group flow. Indeed, as originally shown in \cite{Zamolodchikov1991tba}, from the point of view of statistical mechanics a (1+1)-dimensional QFT at finite temperature $\beta^{-1}$ in infinite volume can be seen equivalently as a zero temperature QFT at finite volume $R$, with the identification $R=\beta$. Therefore, by defining the correlation length $\xi = m^{-1}$, the equivalence of the two quantisation channels yields
\be
r = m\beta = \frac{R}{\xi},
\ee
from which it is clear that the infrared (IR) corresponds to taking $r \to \infty$ and the ultraviolet (UV) limit corresponds to $r \to 0$ (by either keeping $m$ fixed and sending $\beta \to 0$ or by keeping the temperature fixed and sending $m\to 0$). In the UV limit, the theory flows toward a CFT and the scaling function approaches the effective central charge:
\be
\label{scaling function limit Zamolodchikov}
\lim_{r\to 0} c(r) = c_\text{eff}.
\ee
In a GGE of the form $\rho \sim e^{-\sum_s \beta_s Q_s}$, the TBA equation is modified by replacing the driving term $\beta e(\vartheta)$ in \eqref{TBA_eq_thermal} with a linear combination of the one-particle eigenvalues:
\be
w(\vartheta) = \sum_s \beta_s h_s(\vartheta),
\ee
so that 
\be
\label{GGE_TBA_general_definition}
\varepsilon_w(\vartheta) = \sum_s \beta_s h_s(\vartheta) - \varphi*L_w(\vartheta), \quad  L_w(\vartheta) = \log \l 1+e^{-\varepsilon_w(\vartheta)} \r.
\ee
The occupation function $n_w(\vartheta)$ is defined as in \eqref{filling_function_thermal_state} but with $\varepsilon(\vartheta)$ replaced by $\varepsilon_w(\vartheta)$. From the definition of the one-particle eigenvalues \eqref{eigenvalues_even_odd_definition}, it follows that the mass dimensions of the generalised thermodynamic potentials are $[\beta_s]= -s$. In light of this, in the following we take all the non-vanishing $\beta_s$ to be $\beta_s = \beta^s = T^{-s}$. In a spin-$s$ state, only the charge $Q_s$ is coupled to a non-vanishing potential. The corresponding TBA equation is:
\be
\label{spin-s_TBA_eq}
\varepsilon_s(\vartheta) = \beta^s m^s \cosh(s\vartheta)  - \varphi*L_s(\vartheta).
\ee
With this notation, the thermal pseudoenergy is $\varepsilon(\vartheta) \equiv \varepsilon_1 (\vartheta)$ and analogously $L(\vartheta) \equiv L_1 (\vartheta)$,  $n(\vartheta) \equiv n_1 (\vartheta)$.

\begin{figure}[ht]
        \centering
        {\includegraphics[width=\textwidth]{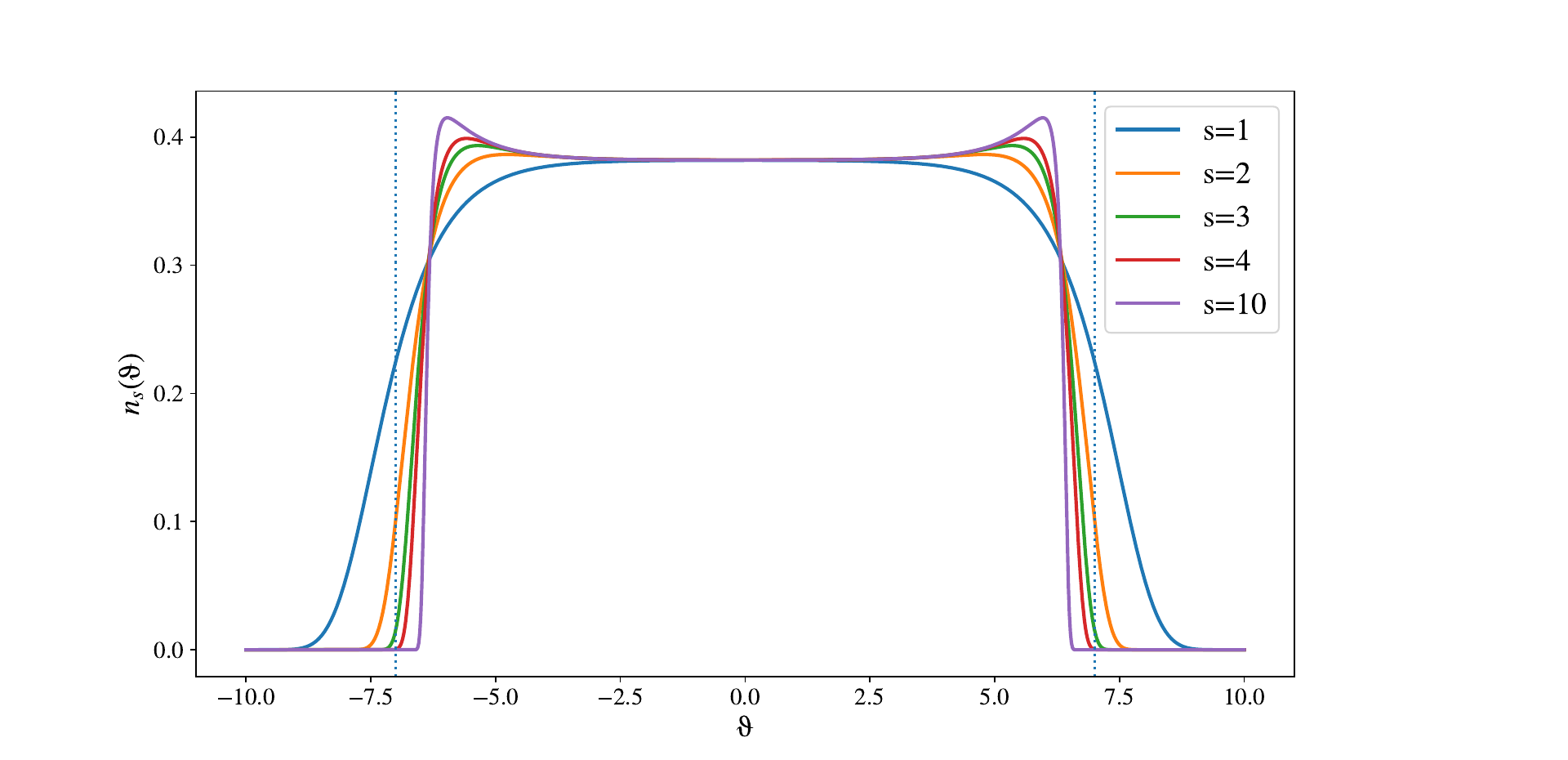}}
        \caption{Occupation functions $n_s(\vartheta)$ for different values of spin $s$ and $r=2e^{-6}$ in the Lee-Yang model. The dashed lines are at $\vartheta = \pm \log\frac{2}{r}$. The Lee-Yang kernel is presented in Section~\ref{Numerical results}.}
    \label{fig:equilibrium_occupation_function}
\end{figure}
We now consider the UV limit of the TBA equation. As $r \to 0$, typically the functions $n(\vartheta)$ and $L(\vartheta)$ display a plateau between $-x$ and $+x$, with
\be
x = \log \frac{2}{r},
\ee
and have double-exponential decay when $|\vartheta| \gg x$. The same behaviour is found for the functions $n_s(\vartheta)$ and $L_s(\vartheta)$, with steeper kinks at $\pm x$ for larger values of $s$, as depicted in Figure \ref{fig:equilibrium_occupation_function}. Referring to the GGE TBA equation \eqref{GGE_TBA_general_definition}, we define the right- and left-shifted pseudoenergies:
\be
\label{shifted_pseudoenergies_def}
\varepsilon_w^\pm(\vartheta) \equiv \varepsilon_w(\vartheta \pm x),
\ee 
and $L_w^\pm(\vartheta)$, $n_w^\pm(\vartheta)$ are defined analogously. Because as $r\to 0$, $r^s\cosh(s(\vartheta \pm x)) \simeq 2^{s-1}e^{\pm s\vartheta}$, in the conformal limit the shifted TBA equations for a spin-$s$ state become (up to exponentially decreasing corrections)
\be
\label{shifted_spin_s_TBA_eq}
\varepsilon_s^\pm(\vartheta) = 2^{s-1} e^{\pm s\vartheta} - \varphi*L^{\pm}_s(\vartheta).
\ee
In particular, in a thermal state $(s=1)$ we obtain the familiar equations for the kink pseudoenergy \cite{Zamolodchikov1991tba}:
\be
\label{TBA_kink_eq}
\varepsilon^\pm (\vartheta) = e^{\pm \vartheta} - \varphi*L^\pm(\vartheta).
\ee
By using the fact that 
\be
\label{L_derivative}
L'_w(\vartheta) = -\varepsilon_w'(\vartheta)n_w(\vartheta),
\ee 
where the prime denotes differentiation w.r.t. $\vartheta$, from \eqref{shifted_spin_s_TBA_eq} we get:
\be
\label{shifted_spin_s_TBA_eq_derivative}
(\varepsilon_s^{\pm})'(\vartheta) = \pm s 2^{s-1} e^{\pm s\vartheta} +\varphi*(n_s^\pm (\varepsilon_s^{\pm})')(\vartheta).
\ee
We conclude with an important remark: changing variables to $\tilde{\vartheta} = \vartheta + x$ (or equivalently $\tilde{\vartheta} = \vartheta - x$), the kink equations \eqref{TBA_kink_eq} coincide with the TBA equations for the left- and right-mover of a CFT in the BLZ formulation \cite{bazhanov1996integrable}:
\be
\label{BLZ_TBA}
\varepsilon_{LM}(\tilde\vartheta) = \frac{\beta m}{2}e^{\tilde\vartheta} - \varphi *L_{LM} (\tilde\vartheta), \quad \varepsilon_{RM}(\tilde\vartheta) = \varepsilon_{LM}(-\tilde\vartheta),
\ee
which describe the properties at temperature $\beta^{-1}$ (or at volume $\beta$) of a theory of two massless particles with energies $e_{LM}(\vartheta)=\frac{m}{2}e^\vartheta$, $e_{RM}(\vartheta)=\frac{m}{2}e^{-\vartheta}$ (see also \cite{Zamolodchikov:1991vx} for the TBA description of massless flows). Notice that in a CFT the quantity $m$ refers to a set energy scale of the theory rather than a mass. The difference between equations \eqref{TBA_kink_eq} and \eqref{BLZ_TBA} is that the former is only valid asymptotically for $m\beta \to 0$, while the latter holds for every finite value of $m\beta$. This means that the expressions that we obtain from an IQFT at the leading order in $r$ when $r\to 0$ coincide with the exact CFT expressions valid at any value of $r$.

\subsection{Currents and Densities in GHD}
The TBA provides a natural quasi-particle picture in which the averages of the conserved quantities are expressed as integrals of the corresponding eigenvalues over the density of occupied states (sometimes also called spectral density) $\rho_p(\vartheta)$:
\be
\label{charge_average_definition_with_state_density}
\text{q}_s \equiv\langle q_s \rangle = \Tr[ q_s \rho_{GGE}] = \int \mathrm{d}\vartheta h_s(\vartheta) \rho_p(\vartheta).
\ee
If the theory is interacting, the charge eigenvalues are ``dressed"  due the presence of scattering. The dressing of a function $h(\vartheta)$ is a map $h(\vartheta) \mapsto h^\text{dr}(\vartheta)$, where $h^\text{dr}(\vartheta)$ satisfies the integral equation:
\be
\label{dressing_definition}
h^\text{dr}(\vartheta) = h(\vartheta) + \varphi * (n h^\text{dr})(\vartheta),
\ee
with $n(\vartheta)$ the TBA occupation function. In the following Section we will make extensive use of two properties of the dressing operation. It is linear:
\be
[\alpha f(\vartheta) + \beta g(\vartheta)]^\text{dr} = \alpha f^\text{dr}(\vartheta) + \beta g^\text{dr}(\vartheta),
\ee
and it is symmetric:
\be
\label{dressing_symmetry_property}
\int \mathrm{d}\vartheta f(\vartheta) n(\vartheta) g^\text{dr}(\vartheta) = \int \mathrm{d}\vartheta f^\text{dr}(\vartheta) n(\vartheta) g(\vartheta).
\ee
Using the definition of the dressed charge eigenvalues and the Bethe constraint between $\rho_p(\vartheta)$ and $n(\vartheta)$ one can rewrite the average $\text{q}_s$ in a more convenient fashion \cite{castro2016emergent}:
\be
\label{average_charge_density_definition_occupation_function}
\text{q}_s = \int \frac{\mathrm{d} p}{2\pi} n (\vartheta) h_s^\text{dr}(\vartheta) = \int \frac{\mathrm{d} \vartheta}{2\pi} e(\vartheta) n (\vartheta) h_s^\text{dr}(\vartheta),
\ee
and analogously it can be shown for the corresponding current $\text{j}_s \equiv\langle j_s \rangle$  that
\be
\label{currents_definition_mirror}
\text{j}_s = \int \frac{\mathrm{d} e}{2\pi} n (\vartheta) h_s^\text{dr}(\vartheta) = \int \frac{\mathrm{d} \vartheta}{2\pi} p(\vartheta) n (\vartheta) h_s^\text{dr}(\vartheta).
\ee
It is possible to write the currents in a way akin to \eqref{charge_average_definition_with_state_density} by introducing the effective velocity of the particles in the theory:
\be
v^\text{eff}(\vartheta) = \frac{(e')^\text{dr}(\vartheta)}{(p')^\text{dr}(\vartheta)} = \frac{p^\text{dr}(\vartheta)}{e^\text{dr}(\vartheta)},
\ee
so that \eqref{currents_definition_mirror} becomes:
\be
\text{j}_s = \int \mathrm{d}\vartheta h_s(\vartheta) v^\text{eff}(\vartheta)\rho_p(\vartheta).
\ee
Let us consider the asymptotics of the dressing equation \eqref{dressing_definition} in the UV limit. By shifting $\vartheta \mapsto \vartheta \pm x$, we get:
\begin{align}
\label{shifted_dressing_equation}
h^{\text{dr},\pm}(\vartheta) &= h(\vartheta \pm x) + \int \frac{\mathrm{d}\gamma}{2\pi}\varphi (\vartheta \pm x -\gamma)n(\gamma)h^\text{dr}(\gamma) \nonumber \\
&= h(\vartheta \pm x) + \int \frac{\mathrm{d}\gamma}{2\pi}\varphi (\vartheta -\gamma)n(\gamma\pm x)h^\text{dr}(\gamma\pm x) \nonumber \\
&= h^\pm(\vartheta) + \varphi* (n^\pm h^{\text{dr},\pm})(\vartheta).
\end{align}
Since for large $x$
\be
h_{s}^\pm(\vartheta) \simeq \frac{m^s}{2}e^{\pm s(\vartheta + x)} = \beta^{-s}2^{s-1} e^{\pm s\vartheta}, \quad h_{-s}^\pm(\vartheta) \simeq \pm\frac{m^s}{2}e^{\pm s(\vartheta + x)} = \pm \beta^{-s}2^{s-1} e^{\pm s\vartheta},
\ee
by comparing equations \eqref{shifted_spin_s_TBA_eq_derivative} and \eqref{shifted_dressing_equation} we obtain the TBA-dressing relations
\be
\label{shifted_dressing_derivative_relation}
h^{\text{dr},\pm}_s(\vartheta) \simeq \beta^{-s}2^{s-1} [e^{\pm s\vartheta}]^\text{dr} \simeq \pm \frac{(\varepsilon_s^{\pm})'(\vartheta)}{s\beta^s},
\ee
asymptotically valid for the even spin-$s$ charge, and
\be
\label{shifted_dressing_derivative_relation_odd}
h^{\text{dr},\pm}_{-s}(\vartheta) \simeq \pm \beta^{-s}2^{s-1} [e^{\pm s\vartheta}]^\text{dr} \simeq \frac{(\varepsilon_s^{\pm})'(\vartheta)}{s\beta^s},
\ee
for the odd spin-$s$ charge. These relations are essential for the derivation of our main results in the next Section.

\subsection{Partitioning Protocol}
Let us now turn to inhomogeneous GGEs. The space-time dependence of the parameters $\beta_s(x,t)$ implies that in the quasi-particle picture also the particle density, the state density and the occupation function depend on $(x,t)$, that is $\rho_p(\vartheta) \mapsto \rho_p(\vartheta; x,t)$, $\rho_s(\vartheta) \mapsto \rho_s(\vartheta;x,t)$ and $n(\vartheta) \mapsto n(\vartheta;x,t)$. All the fundamental GHD equations can be described in terms of the state coordinates $n(\vartheta; x,t)$, and consequently acquire an $(x,t)$-dependence. In particular, the averages of charge and current densities are now inhomogeneous, and Euler's equation \eqref{Euler_equation_fluid_cell} can be rewritten as
\be
\label{Euler_equation_normal_modes}
\partial_t n(\vartheta; x,t) + v^\text{eff}(\vartheta; x,t) \partial_x n(\vartheta; x,t)=0.
\ee
The function $n(\vartheta; x,t)$ thus acquires the meaning of (distribution of) hydrodynamic normal modes, the modes being transported with velocity $v^\text{eff}(\vartheta; x,t) = p^\text{dr}(\vartheta;x,t)/e^\text{dr}(\vartheta;x,t)$. 

Within this picture, the partitioning protocol described in the introduction amounts to solving equation \eqref{Euler_equation_normal_modes} equipped with the initial condition:
\be
n(\vartheta;x,t=0^+) = n_L(\vartheta) \Theta(-x) + n_R(\vartheta) \Theta(x),
\ee
where $n_L(\vartheta)$ and $n_R(\vartheta)$ characterise the reservoirs in the left and right half-lines. This is the Riemann problem of hydrodynamics. Because both the initial conditions and the Euler equation are invariant under the scaling $(x,t) \mapsto (\lambda x, \lambda t)$, one can reformulate the problem in terms of a dimensionless ray $\xi=x/t$:
\be
\label{Riemann_problem_generic}
\begin{cases}
 &\l v^\text{eff}(\vartheta;\xi) -\xi\r \partial_\xi n(\vartheta; \xi)=0, \\
 & \underset{\xi \to \infty}{\lim} n(\vartheta;\xi) = n_R(\vartheta), \quad \underset{\xi \to -\infty}{\lim}n(\vartheta;\xi) = n_L(\vartheta).
\end{cases}
\ee
It can be shown that, because of linear degeneracy of the modes in GHD \cite{castro2016emergent, bertini_ghd, doyon2020lecture}, each mode $n(\vartheta, \xi) $ has a jump discontinuity exactly at the ray $\xi$ corresponding to its velocity. In most cases\footnote{As shown in \cite{staircase} an interesting example of a non-monotonic effective velocity is Zamolodchikov's staircase model \cite{zamolodchikov2006resonance}, for which the solution (\ref{sol}) needs to be generalised to allow for multiple discontinuities.}, $v^\text{eff}$ is a monotonic function of $\vartheta$. In this situation, the solution to \eqref{Riemann_problem_generic} is:
\be
\begin{cases}
&n(\vartheta;\xi)=n_L(\vartheta)\Theta(\vartheta - \vartheta_*(\xi))+n_R(\vartheta)\Theta(\vartheta_*(\xi)-\vartheta), \\
&v^\text{eff}(\vartheta_*(\xi),\xi)=\xi.
\end{cases}
\label{sol}
\ee
Observables such as the expectation values of currents and charge densities continuously vary within the light-cone defined by $|\xi|\le 1$. The NESS is the state at $\xi=0$.
\section{Main Results: NESS Averages of Higher-Spin conserved charges}
\label{Section: main results}
In this Section, we derive the main results of this paper: the UV limit of (averages of) conserved current densities $\text{j}_s$ and charge densities $\text{q}_s$ in the NESS arising after a partitioning protocol. We consider the simple case of a single-particle QFT and two types of asymptotic boundary conditions for the Riemann problem, corresponding to thermal states (with TBA equation \eqref{TBA_eq_thermal}) and spin-$s$ states (with TBA equation \eqref{spin-s_TBA_eq}). In both cases, the results display a power law dependence on the temperatures  of the asymptotic states, and we interpret the coefficients as spin-dependent generalisations of the CFT effective central charge. Finally, we consider the interesting case in which a higher-spin charge is taken as generator of the time evolution. An extension of our results to relativistic IQFTs with multi-particle spectra is given in Appendix \ref{Appendix:Many-particle theories}.
\subsection{Thermal Reservoirs}
We start by looking at the simple case of a homogeneous thermal reservoir, described by the TBA equation \eqref{TBA_eq_thermal}. The average of the charge density $q_s$ is given by \eqref{average_charge_density_definition_occupation_function}. The UV limit is obtained by shifting the rapidity variable and making use of  \eqref{shifted_dressing_derivative_relation}:
\begin{align}
\text{q}_s = \int\frac{\mathrm{d}\vartheta}{2\pi}\,e(\vartheta)n(\vartheta)h_s^\text{dr}(\vartheta) &= \frac{1}{\pi}\int_{0}^{+\infty} \mathrm{d}\vartheta\,e(\vartheta)n(\vartheta)h_s^\text{dr}(\vartheta) \nonumber \\
&= \frac{1}{\pi}\int_{-x}^{+\infty} \mathrm{d}\vartheta\,e^+(\vartheta)n^+(\vartheta)h_s^{\text{dr},+}(\vartheta) \nonumber \\
&\simeq \frac{2^{s-1}}{\pi \beta^{s+1}}\int_{-x}^{+\infty} \mathrm{d}\vartheta\,e^\vartheta n^+(\vartheta)[e^{s\vartheta}]^\text{dr}.
\end{align}
Here and in the following, the notation $\simeq$ means that as $r \to 0$ the relative error goes to zero at least as fast as $\mathcal{O}(r^\alpha)$ for some positive $\alpha$. For $s=1$, i.e. when we consider the average energy density, from $[e^{\vartheta}]^\text{dr} = \varepsilon^\prime(\vartheta)$ and \eqref{L_derivative} the well-known CFT result is immediate, after integration by parts:
\be
\label{Stefan-Boltzmann}
\text{q}_1 = \frac{\pi c_\text{eff}}{6\beta^2},
\ee
where we used the limit expression of the effective central charge:
\begin{align}
c_\text{eff} &= \lim_{r\to 0} \frac{3 r}{\pi^2} \int \mathrm{d}\vartheta \cosh (\vartheta)L(\vartheta) 
\nonumber \\ &= \lim_{x\to \infty} \frac{6}{\pi^2} \int_{-x}^{+\infty}\mathrm{d}\vartheta\,e^\vartheta L^+ (\vartheta) = \lim_{x\to \infty} \frac{6}{\pi^2} \int_{-\infty}^{x}\mathrm{d}\vartheta\,e^{-\vartheta} L^- (\vartheta).
\end{align}
For $s\ne 1$, on the other hand, we can take advantage of the fact that in the limit $x \to \infty$ the integral extends over all $\mathbb{R}$ and ``move" the dressing operation using equation \eqref{dressing_symmetry_property}:
\begin{align}
\label{q_s_equilibrium_computation}
\text{q}_s \simeq \frac{2^{s-1}}{\pi\beta^{s+1}}\int_{-x}^{+\infty} \mathrm{d}\vartheta\,[e^\vartheta]^\text{dr} n^+(\vartheta)e^{s\vartheta} &= -\frac{2^{s-1}}{\pi\beta^{s+1}}\int_{-x}^{+\infty} \mathrm{d}\vartheta\,(L^{+})'(\vartheta)e^{s\vartheta} \nonumber \\
&= \frac{s2^{s-1} }{\pi\beta^{s+1}}\int_{-x}^{+\infty} \mathrm{d}\vartheta\,L^+(\vartheta)e^{s\vartheta}\nonumber \\&= \frac{s2^{s-1} \pi}{6\beta^{s+1}} \mathcal{C}(s),
\end{align}
where again the second line is obtained integrating by parts\footnote{The boundary term, evaluated for large but finite $x$, gives a subleading correction which is always of order $\mathcal{O}(\beta^{-1})$. Indeed, $\frac{1}{\beta^{s+1}}\left[L(\vartheta + x)e^{s\vartheta}\right]^{\infty}_{-x} = -\frac{L(0)}{\beta^{s+1}}e^{-s x} = - \frac{L(0)}{\beta^{s+1}}\l \frac{m \beta}{2}\r^s$. The term at $\vartheta \to \infty$ vanishes because of the double exponential decay of $L$ at large rapidities.}. The function $\mathcal{C}(s)$ provides a generalisation of the effective central charge in the sense that:
\begin{align}
\label{C(s)_definition}
\mathcal{C}(s) &\equiv \lim_{r\to 0} \frac{6 r^s}{2^{s}\pi^2} \int \mathrm{d}\vartheta \cosh (s\vartheta)L(\vartheta) \nonumber \\
&= \lim_{x\to \infty} \frac{6}{\pi^2} \int_{-x}^{+\infty}\mathrm{d}\vartheta\,e^{s\vartheta} L^+ (\vartheta) = \lim_{x\to \infty} \frac{6}{\pi^2} \int_{-\infty}^{x}\mathrm{d}\vartheta\,e^{-s\vartheta} L^- (\vartheta),
\end{align}
normalised in such a way that $\mathcal{C}(1)=c_\text{eff}$. In Section \ref{Exact results for the massive free fermion} we show how an explicit formula for $\mathcal{C}(s)$ in terms of polylogarithms may be obtained for the free fermion, while a numerical evaluation of this function for different theories is presented in Section \ref{Numerical results}. 

The result \eqref{Stefan-Boltzmann}, first derived in \cite{blote1986conformal,affleck1988universal}, was identified in \cite{cardy2010ubiquitous} as the (1+1)-dimensional analogue of  Stefan-Boltzmann's law. In the same spirit, we can interpret the equilibrium result \eqref{q_s_equilibrium_computation} as a generalisation of that law in which the pressure of radiation is not related to the energy density but to the density of a higher-spin charge $Q_s$. Notice that, for symmetry reasons, at equilibrium the current averages are $\text{j}_s=0$ and $\text{j}_{-s}=\text{q}_s$. 

Out of equilibrium, in the NESS $\xi=0$, the occupation function $n(\vartheta)$ is (assuming the monotonicity of $v^\text{eff}(\vartheta_*,0)$):
\be
\label{occupation_function_NESS}
n(\vartheta)=n_L(\vartheta)\Theta(\vartheta - \vartheta_*)+n_R(\vartheta)\Theta(\vartheta_*-\vartheta), \quad
v^\text{eff}(\vartheta_*,0)=0.
\ee
The occupation functions $n_{L/R}(\vartheta)$ are defined by \eqref{filling_function_thermal_state} with $\varepsilon(\vartheta) \rightarrow \varepsilon_{L/R}(\vartheta)$, and $\varepsilon_{L/R}(\vartheta)$ are in turn given by \eqref{TBA_eq_thermal} with $\beta \rightarrow \beta_{L/R}$. The only solution to $v^\text{eff}(\vartheta_*, 0)= 0$  is in first approximation given by $\vartheta_* =0$. This is exact at equilibrium, as the numerator $p^\text{dr}(\vartheta) = \beta^{-1} \varepsilon'(\vartheta)$ of the effective velocity vanishes at the central point of the symmetric plateau. Out of equilibrium, the position of the discontinuity depends on the two temperatures, but the numerical results show that it changes extremely slowly as the latter vary, and in Appendix \ref{appendix:Useful bounds on the effective velocity} we prove that $\vartheta_*$ is always well within the plateau:
\be
-x_R \ll \vartheta_* \ll x_L, \quad x_R = \ln \frac{2}{r_R}, \quad x_L= \ln \frac{2}{r_L}.
\ee
Equation \eqref{occupation_function_NESS} implies that the expression for the average of the charge density $q_s$ in the NESS, for which we use the same symbol $\text{q}_s$, splits into two integrals which correspond to the contributions of quasi-particles coming from the two thermal baths:
 \begin{align}
    \text{q}_s &=\int_{-\infty}^{\vartheta_*} \frac{\mathrm{d}\vartheta}{2\pi}\,e(\vartheta)n_R(\vartheta)h_s^\text{dr}(\vartheta) + \int_{\vartheta_*}^{\infty}\frac{\mathrm{d}\vartheta}{2\pi}\,e(\vartheta)n_L(\vartheta)h_s^\text{dr}(\vartheta) \nonumber\\
    &= \int_{-\infty}^{\vartheta_* + x_R} \frac{\mathrm{d}\vartheta}{2\pi}\,e^-(\vartheta)n^-_R(\vartheta)h_s^{\text{dr},-}(\vartheta) + \int_{\vartheta_*-x_L}^{\infty}\frac{\mathrm{d}\vartheta}{2\pi}\,e^+(\vartheta)n^+_R(\vartheta)h_s^{\text{dr},+}(\vartheta) \nonumber\\
    &\simeq \frac{2^{s-1}}{2\pi\beta_R^{s+1}}\int_{-\infty}^{x_R} \mathrm{d}\vartheta\,e^{-\vartheta}n_R^-(\vartheta)[e^{-s\vartheta}]^\text{dr} + \frac{2^{s-1}}{2\pi\beta_L^{s+1}}\int_{-x_L}^{\infty} \mathrm{d}\vartheta\,e^{\vartheta}n_L^+(\vartheta)[e^{s\vartheta}]^\text{dr}.
\end{align}
As in the equilibrium case, one can now move the dressing from $[e^{\pm s\vartheta}]^\text{dr}$ to $e^{\pm\vartheta}$ in both the integrals and then repeat the same steps of \eqref{q_s_equilibrium_computation}. There is, however, a subtlety: the functions $[e^{s\vartheta}]^\text{dr}$ and $[e^{-s\vartheta}]^\text{dr}$ are dressed with the global occupation function \eqref{occupation_function_NESS}, but are integrated against $n_L^+(\vartheta)$ and $n_R^-(\vartheta)$ respectively. Therefore, rigorously speaking, the symmetry property \eqref{dressing_symmetry_property} does not apply. Nonetheless, in the high temperature regime (obtained by taking both $x_R$, $x_L \gg 1$) the plateau values of $n_L(\vartheta)$ and that of $n_R(\vartheta)$ coincide, which justifies the use of \eqref{dressing_symmetry_property} in the UV limit. Therefore, after moving the dressing operation we have:
\begin{align}
\label{q_s NESS}
\text{q}_s &\simeq \frac{2^{s-1}}{2\pi\beta_R^{s+1}}\int_{-\infty}^{ x_R} \mathrm{d}\vartheta\,[e^{-\vartheta}]^\text{dr}n_R^-(\vartheta)e^{-s\vartheta} + \frac{2^{s-1}}{2\pi\beta_L^{s+1}}\int_{-x_L}^{\infty} \mathrm{d}\vartheta\,[e^{\vartheta}]^\text{dr}n_L^+(\vartheta)e^{s\vartheta} \nonumber \\
    &=-\frac{2^{s-1}}{2\pi\beta_R^{s+1}}\int_{-\infty}^{ x_R} \mathrm{d}\vartheta\,(\varep^-_R)'(\vartheta)n_R^-(\vartheta)e^{-s\vartheta} + \frac{2^{s-1}}{2\pi\beta_L^{s+1}}\int_{-x_L}^{\infty} \mathrm{d}\vartheta\,(\varep^+_L)'(\vartheta)n_L^+(\vartheta)e^{s\vartheta} \nonumber \\
    &=\frac{s 2^{s-1}}{2\pi\beta_R^{s+1}}\int_{-\infty}^{ x_R} \mathrm{d}\vartheta\,L_R^-(\vartheta)e^{-s\vartheta} + \frac{s 2^{s-1}}{2\pi\beta_L^{s+1}}\int_{-x_L}^{\infty} \mathrm{d}\vartheta\,L_L^+(\vartheta)e^{s\vartheta} \nonumber \\
    &=\frac{s 2^{s-1}\pi}{12}\mathcal{C}(s)\l\frac{1}{\beta_L^{s+1}}+\frac{1}{\beta_R^{s+1}}\r.
 \end{align}
For the average current density:
\be
\label{average thermal current split}
 \text{j}_s =\int_{-\infty}^{\vartheta_*} \frac{\mathrm{d}\vartheta}{2\pi}\,p(\vartheta)n_R(\vartheta)h_s^\text{dr}(\vartheta) + \int_{\vartheta_*}^{\infty}\frac{\mathrm{d}\vartheta}{2\pi}\,p(\vartheta)n_L(\vartheta)h_s^\text{dr}(\vartheta),
\ee
the calculation proceeds exactly in the same way, the only difference coming from an extra minus sign in front of the first integral when taking the large temperature limit after the rapidity shift. Therefore, in the conformal limit:
\beq
\label{j_s NESS}
 \text{j}_s = \frac{s 2^{s-1}\pi}{12}\mathcal{C}(s)\l\frac{1}{\beta_L^{s+1}}-\frac{1}{\beta_R^{s+1}}\r.
\eeq
For $s=1$, the expressions reduce to \eqref{CFT_energy_results} as expected. A computation of the NESS average energy current density $\text{j}_1$ at finite temperatures was already present in \cite{castro2014thermodynamic, horvath2019hydrodynamics}, where the notion of dynamical central charge was defined. 

The computations for the odd spin-$s$ charges are carried out using the asymptotic relation \eqref{shifted_dressing_derivative_relation_odd} and present no differences with respect to the previous case, yielding:
\be
\text{q}_{-s} = \text{j}_s, \quad \text{j}_{-s} = \text{q}_s.
\ee
For a generic value of the spin, the equation above is exact only in the conformal limit. For $s=1$, however, the relation $\text{q}_{-1}=\text{j}_{1}$, which is nothing but the statement that the (average) momentum density equates the (average) energy current density, is valid at every value of the mass and temperature. We conclude with a remark: the large (positive/negative) rapidity asymptotics of $h_s(\vartheta)$, that is the functions  $\frac{m^s}{2}e^{\pm s \vartheta}$, are the charge eigenvalues of the CFT right- ($+$) and left- ($-$) movers. Hence, in the conformal limit the only contribution to the thermal average in the NESS coming from the left (right) reservoir is that of the right- (left-) movers, in agreement with the findings of \cite{bernard2012energy, Bernard_2014, bernard2016conformal}. 

\subsection{Non-Thermal Reservoirs and Generalised Times}
We now discuss the case in which the asymptotic states of the system at $x \to \pm \infty$ are GGEs. Although one should in principle consider states described by infinitely many thermodynamic potentials, the situation in which it is possible to extract analytic results from the TBA formalism is that in which there is a single non-vanishing potential $\beta_s = \beta^s$, that is, the reservoirs are in spin-$s$ states.  

Let us start again by considering the equilibrium situation, with $n_s(\vartheta)= (1+e^{\varep_s(\vartheta)})^{-1}$, $\varep_s(\vartheta)$ solution of \eqref{spin-s_TBA_eq}. The average $\text{q}_s^{(s)}$ of the conserved density $q_s$ in the UV limit is\footnote{In interacting IQFTs, exact asymptotic formulae can be obtained only for the special case when the spin of the average charge density is the same as that of the state. In Section \ref{Exact results for the massive free fermion} we provide exact results for the free fermion also when the two spins are different.}: 
\begin{align}
\label{q_s^s_equilibrium_computation}
\text{q}_s^{(s)} = \int\frac{\mathrm{d}\vartheta}{2\pi}\,e(\vartheta)n_s(\vartheta)h_s^\text{dr}(\vartheta) 
&= \frac{1}{\pi}\int_{-x}^{+\infty} \mathrm{d}\vartheta\,e^+(\vartheta)n^+_s(\vartheta)h_s^{\text{dr},+}(\vartheta) \nonumber \\
& \simeq \frac{1}{s \pi \beta^{s+1}}\int_{-x}^{+\infty} \mathrm{d}\vartheta\,e^\vartheta n_s^+(\vartheta)(\varep^+_s)'(\vartheta) \nonumber \\
& = \frac{\pi}{ 6 s \beta^{s+1}}\widetilde{\mathcal{C}}(s),
\end{align}
with
\begin{align}
\label{Tilde_C(s)_definition}
\widetilde{\mathcal{C}}(s) &\equiv \lim_{r\to 0} \frac{3 r}{\pi^2} \int \mathrm{d}\vartheta \cosh (\vartheta)L_s(\vartheta) \nonumber \\
&= \lim_{x\to \infty} \frac{6}{\pi^2} \int_{-x}^{+\infty}\mathrm{d}\vartheta\,e^{\vartheta} L^+_s (\vartheta) = \lim_{x\to \infty} \frac{6}{\pi^2} \int_{-\infty}^{x}\mathrm{d}\vartheta\,e^{-\vartheta} L^-_s (\vartheta).
\end{align}
The average currents and densities in the NESS are obtained in a straightforward way, by following the procedure outlined in the previous Section, giving:
\be
\label{q^s_s e j^s_s NESS}
 \text{q}_s^{(s)} = \frac{\pi}{ 12 s }\widetilde{\mathcal{C}}(s)\l\frac{1}{\beta_L^{s+1}}+\frac{1}{\beta_R^{s+1}}\r, \quad \text{j}_s^{(s)} = \frac{\pi}{ 12 s }\widetilde{\mathcal{C}}(s)\l\frac{1}{\beta_L^{s+1}}-\frac{1}{\beta_R^{s+1}}\r,
\ee
and the odd spin-$s$ averages are similarly given by $\text{q}_{-s}^{(s)} = \text{j}_s^{(s)}$, $\text{j}_{-s}^{(s)} = \text{q}_s^{(s)}$.

Comparing \eqref{q_s_equilibrium_computation} and \eqref{q_s^s_equilibrium_computation} we see that the average of a spin-$s$ charge density, in the high temperature regime, is proportional to $1/\beta^{s+1}=T^{s+1}$ both when the system is in a thermal state or in a spin-$s$ state. In the next Section we show that the same scaling law is obtained for the free fermion also in the case of a spin-$s$ state and a spin-$k$ conserved charge, with $s \ne k$ (see Subsection \ref{InGGE}). In that case, the scaling still goes with powers of $\beta_L, \beta_R$ where the power is $k+1$, that is, the power law is dictated by the spin of the conserved quantity rather than the spin of the state, although both are involved non-trivially in the proportionality coefficient. This dependence on the temperature follows from simple dimensional arguments and is ultimately due to the specific choice of the thermodynamic potentials $\beta_s = \beta^s$. The coefficients, however, are generally not the same in the two states. Although it is not immediately evident in the case of interacting theories, the functions $\mathcal{C}(s)$ and $\widetilde{\mathcal{C}}(s)$ depend on the spin $s$ in rather different ways. In particular, in a free fermion theory $\mathcal{C}(s)$ grows faster than an exponential while $\widetilde{\mathcal{C}}(s)$ decreases. Moreover, even though the integrals are normalised in such a way that $\mathcal{C}(1)=\widetilde{\mathcal{C}}(1)=c_\text{eff}$, for generic $s\ne 1$ they cannot be related to the effective central charge via the standard dilogarithm identities \cite{Zamolodchikov1991tba, Klassen1991}.

Surprisingly, a physical setup in which $c_{\rm eff}$ appears again as a proportionality coefficient in the UV expression of the steady state currents is that in which the time evolution itself is governed by a higher-spin charge. Indeed, because there are several conserved charges in involution, one can take a charge $Q_k$ as the generator of the time evolution and define a generalised time variable $t_k$ through:
\be
\partial_{t_k} \mathcal{O} \equiv i[Q_k, \mathcal{O}],
\ee
for any local observable $\mathcal{O}$. If $k=1$ the charge is the Hamiltonian and $t_1 \equiv t$ is the usual time. But, as discussed in \cite{doyon2017note}, it is also possible to define generalised current densities $j_{s,k}$ through the operator equation: 
\be
\partial_{t_k} q_s + \partial_x j_{s,k}=0\,.
\label{tk}
\ee
In the hydrodynamic approximation, \eqref{tk} applies as usual also to the averages $\texttt{q}_s$ and $\texttt{j}_{s,k}$. While the expectation value of the charge density is not affected by the choice of a different time parameter, the GHD expression of $\text{j}_{s,k}$ (which is fixed by the continuity equation up to a constant) is \cite{doyon2017note,yoshimura2020collision,borsi2020current,borsi2021current,bajnok2020exact}:
\be
\text{j}_{s,k} = \int {\texttt d}\vartheta \, v^\text{eff}[h_k](\vartheta) \rho_p(\vartheta) h_s (\vartheta) = \int \frac{\mathrm{d}\vartheta}{2\pi} (h_k^\prime)^\text{dr}(\vartheta) n(\vartheta) h_s(\vartheta),
\ee
where the generalised effective velocity is defined by:
\be
v^\text{eff}[h](\vartheta) \equiv \frac{(h^\prime)^\text{dr}(\vartheta)}{(p^\prime)^\text{dr}(\vartheta)},
\ee
so that when $h(\vartheta)=e(\vartheta)$, then $v^\text{eff}[e](\vartheta) =  v^\text{eff}(\vartheta)$. In terms of the occupation function $n(\vartheta)$, the GHD equation for the flow generated by $Q_k$ becomes:
\be
\partial_{t_k} n(\vartheta) + v^\text{eff}[h_k](\vartheta) \partial_x n(\vartheta).
\ee
Because the equation is scale-invariant, it can be recast in terms of a new dimensionless ray $\xi_k = \frac{x}{t_k}$ and the Riemann problem has the same form \eqref{Riemann_problem_generic}, with the NESS now being the state at $\xi_k=0$. The effective central charge emerges in the computation of the NESS average $\text{j}_{s,s}$, that is, when the charge ruling the time evolution is the same appearing in the continuity equation through its density. Moreover, suppose that the system is in a spin-$s$ state, so that $Q_s$ is also the only charge in the GGE with a non vanishing generalised potential $\beta^s$. Then the NESS average, denoted by $\text{j}^{(s)}_{s,s}$ is:
\begin{align}
\label{j^s_s_s NESS}
 \text{j}^{(s)}_{s,s} &= \int \frac{\mathrm{d}\vartheta}{2\pi} h^\prime_s(\vartheta) n_s(\vartheta) h_s^\text{dr}(\vartheta) \nonumber \\
 &= \int_{-\infty}^{\vartheta_* + x_R} \frac{\mathrm{d}\vartheta}{2\pi} (h^{-}_s)'(\vartheta) n_{s,R}^-(\vartheta) h_s^{\text{dr},-}(\vartheta) + \int_{\vartheta_* - x_L}^{\infty} \frac{\mathrm{d}\vartheta}{2\pi} (h^{+}_s)'(\vartheta) n_{s,L}^+(\vartheta) h_s^{\text{dr},+}(\vartheta) \nonumber \\
 &\simeq -\frac{s 2^s}{4\pi  \beta_R^{s}}\int_{-\infty}^{ x_R} \mathrm{d}\vartheta\,e^{-s\vartheta}n_{s,R}^-(\vartheta)\l-\frac{(\varep_s^{-})'(\vartheta)}{s\beta_R^s}\r + \frac{s 2^s }{4\pi \beta_L^{s}}\int_{-x_L}^{ \infty} \mathrm{d}\vartheta\,e^{s\vartheta}n_{s,L}^+(\vartheta)\l\frac{(\varep_s^{+})'(\vartheta)}{s\beta_L^s}\r \nonumber \\
    &= -\frac{2^s}{4\pi \beta_R^{2s}}\int_{-\infty}^{x_R}\mathrm{d}\vartheta e^{-s\vartheta} (L_{s,R}^{-})'(\vartheta) - \frac{2^s}{4\pi \beta_L^{2s}}\int_{-x_L}^{\infty}\mathrm{d}\vartheta e^{s\vartheta} (L_{s,L}^{+})'(\vartheta) \nonumber \\
    &= -\frac{s 2^s}{4\pi \beta_R^{2s}}\int_{-\infty}^{x_R}\mathrm{d}\vartheta e^{-s\vartheta} L_{s,R}^{-}(\vartheta) + \frac{s 2^s}{4\pi \beta_L^{2s}}\int_{-x_L}^{\infty}\mathrm{d}\vartheta e^{s\vartheta} L_{s,L}^{ +}(\vartheta) \nonumber \\
    &= \frac{\pi c_\text{eff}}{12}\l \frac{1}{\beta_L^{2s}} - \frac{1}{\beta_R^{2s}}\r.   
\end{align}
The fifth line is obtained integrating by parts and in the last line we used the fact that the generalised spin-$s$ scaling function
\be
\label{spin-s scaling function def}
c_s(r) \equiv \frac{3 s r^s}{\pi^2} \int \mathrm{d}\vartheta L_s(\vartheta) \cosh(s\vartheta), 
\ee
reproduces the effective central charge in the UV limit:
\be
\label{spin-s scaling function limit}
\lim_{r\to 0} c_s(r) = \lim_{x\to \infty} \frac{6 s 2^{s-1}}{\pi^2}\int_{-\infty}^{x}\mathrm{d}\vartheta L_s^-(\vartheta) e^{-s\vartheta} = \lim_{x\to \infty} \frac{6 s 2^{s-1}}{\pi^2}\int_{-x}^{\infty}\mathrm{d}\vartheta L_s^+(\vartheta) e^{s\vartheta} = c_\text{eff}.
\ee
A proof of this statement is given in Appendix \ref{appendix:Spin-dependent scaling function}. Equation \eqref{j^s_s_s NESS} displays a dependence on $\beta_L$, $\beta_R$ different from that of \eqref{j_s NESS} and \eqref{q^s_s e j^s_s NESS}. Intuitively, the physical reason for this is that the time parameter $t_s$, being associated to a higher-spin charge, scales differently than the Hamiltonian time $t_1$ as the fundamental length of the system is varied. 

\section{Exact Results for the Massive Free Fermion}
\label{Exact results for the massive free fermion}
In a free fermion theory, the lack of interactions allows one to obtain exact expressions for the expectation values of any local charge in a partitioning protocol. These results are valid without any approximation at all values of the temperature or, in other words, when the theory is genuinely far from the conformal point. By taking the UV limit $r\to 0$ we can then compare the free fermion expectation values with the more general results derived in the previous Section. The availability of exact expressions for the massive free fermion is a consequence of the fact that the dressing of the charge eigenvalues is trivial in this case, $h^\text{dr}(\vartheta)=h(\vartheta)$, and the effective velocity:
\be
\label{ff_effective_velocity}
v^\text{eff}(\vartheta) = \tanh \vartheta,
\ee
has a zero at $\vartheta_* = 0$. Hence, the NESS occupation function is:
\be
\label{ff_ness_general}
n(\vartheta) = n_L(\vartheta)\Theta(\vartheta)+  n_R(\vartheta)\Theta(-\vartheta)\,.
\ee
In this Section, we derive the expressions for the free fermion energy, momentum and higher-spin charge and current averages in the partitioning protocol. As done in the previous Section, we first consider asymptotic thermal reservoirs and then asymptotic spin-$s$ states. We compute the coefficients $\mathcal{C}(s)$, $\widetilde{\mathcal{C}}(s)$ and relate these to the effective central charge.
\subsection{Inhomogeneous Thermal State}
The TBA equations for the two reservoirs are simply
\be
\varepsilon_{R/L}(\vartheta) = m \beta_{R/L} \cosh{\vartheta} \equiv r_{R/L}\cosh{\vartheta}\,,
\ee
and \eqref{ff_ness_general} becomes:
\be
n(\vartheta) = \frac{1}{1+ e^{r_L \cosh{\vartheta}}}\Theta(\vartheta)+  \frac{1}{1+ e^{r_R \cosh{\vartheta}}}\Theta(-\vartheta).
\ee
It follows that the energy and momentum averages are given by
\begin{align}
\text{q}_1 &= \int \frac{\mathrm{d}\vartheta}{2\pi} n(\vartheta) e^2(\vartheta) = \frac{m^2}{2\pi} \l \int_0^{\infty} \mathrm{d}\vartheta \frac{\cosh^2{\vartheta}}{1+ e^{r_L \cosh{\vartheta}}} +  \int_0^{\infty} \mathrm{d}\vartheta \frac{\cosh^2{\vartheta}}{1+ e^{r_R \cosh{\vartheta}}}\r, \\
\text{q}_{-1}=\text{j}_{1} &= \int \frac{\mathrm{d}\vartheta}{2\pi} n(\vartheta) e(\vartheta)p(\vartheta) = \frac{m^2}{2\pi} \l \int_0^{\infty} \mathrm{d}\vartheta \frac{\cosh{\vartheta}\sinh{\vartheta}}{1+ e^{r_L \cosh{\vartheta}}} -  \int_0^{\infty} \mathrm{d}\vartheta \frac{\cosh{\vartheta}\sinh{\vartheta}}{1+ e^{r_R \cosh{\vartheta}}}\r, \\
\text{j}_{-1} &= \int \frac{\mathrm{d}\vartheta}{2\pi} n(\vartheta) p^2(\vartheta) = \frac{m^2}{2\pi} \l \int_0^{\infty} \mathrm{d}\vartheta \frac{\sinh^2{\vartheta}}{1+ e^{r_L \cosh{\vartheta}}} +  \int_0^{\infty} \mathrm{d}\vartheta \frac{\sinh^2{\vartheta}}{1+ e^{r_R \cosh{\vartheta}}}\r. 
\end{align}
It is useful at this point to introduce the integrals ($\alpha, \beta \in \mathbb{R}$, $z \in \mathbb{R}_+$):
\begin{subequations}
\label{ff_integrals_definition}
\begin{align}
\mathcal{I}_{\alpha, \beta}^{++}(z) &\equiv \int_0^{\infty} \mathrm{d}\vartheta \frac{\cosh{(\alpha \vartheta)}\cosh{(\beta\vartheta)}}{1+ e^{z \cosh{\vartheta}}} ,\\
\mathcal{I}_{\alpha, \beta}^{+-}(z) &\equiv \int_0^{\infty} \mathrm{d}\vartheta \frac{\cosh{(\alpha \vartheta)}\sinh{(\beta\vartheta)}}{1+ e^{z \cosh{\vartheta}}} , \quad \mathcal{I}_{\alpha, \beta}^{-+}(z) \equiv \mathcal{I}_{\beta, \alpha}^{+-}(z),\\
\mathcal{I}_{\alpha, \beta}^{--}(z) &\equiv \int_0^{\infty} \mathrm{d}\vartheta \frac{\sinh{(\alpha \vartheta)}\sinh{(\beta\vartheta)}}{1+ e^{z \cosh{\vartheta}}} .
\end{align}
\end{subequations}
We compute the closed-form analytic expressions of these integrals (where they exist) and their asymptotic expansions as $z \to 0$ in Appendix \ref{Appendix:Free fermion integrals and finite temperature corrections}. Using those results, we obtain:
\begin{align}
\text{q}_1 &= \frac{m^2}{2\pi} [\mathcal{I}_{1,1}^{++}(r_L)+\mathcal{I}_{1,1}^{++}(r_R)] \nonumber \\&= \frac{m^2}{4\pi} \sum_{n=1}^{\infty}(-1)^{n+1}[K_2(nr_L) + K_0(nr_L) + K_2(nr_R) + K_0(nr_R)], \label{ff_energy_density}\\
\text{j}_{-1} &= \frac{m^2}{2\pi} [\mathcal{I}_{1,1}^{--}(r_L)+\mathcal{I}_{1,1}^{--}(r_R)) \nonumber \\ &= \frac{m^2}{4\pi} \sum_{n=1}^{\infty}(-1)^{n+1}[K_2(nr_L) - K_0(nr_L) + K_2(nr_R) - K_0(nr_R)], \label{ff_momentum_current}\\
\text{q}_{-1} &= \text{j}_1 =  \frac{m^2}{2\pi} [\mathcal{I}_{1,1}^{+-}(r_L)-\mathcal{I}_{1,1}^{+-}(r_R)] \nonumber \\
&= \frac{m^2}{4\pi} \left[\frac{\ln(1+e^{-r_L})}{r_L} - \frac{\text{Li}_2(-e^{-r_L})}{r_L^2} - \frac{\ln(1+e^{-r_R})}{r_R} + \frac{\text{Li}_2(-e^{-r_R})}{r_R^2}\right], \label{ff_energy_current}
\end{align} 
where we used the integral representation of the modified Bessel function of the second kind
\be
\label{modified_bessel_definition}
K_{\nu}(z) = \int_0^{\infty} \mathrm{d}\vartheta \cosh{(\nu \vartheta)} e^{-z \cosh{\vartheta}}\,,
\ee
and
\be
\text{Li}_2(z) \equiv \sum_{n=1}^{\infty}\frac{z^n}{n^2},
\ee
is the dilogarithm function. We stress that expressions \eqref{ff_energy_density}, \eqref{ff_momentum_current}, \eqref{ff_energy_current} are exact and valid at any $r_L$, $r_R > 0$. By making use of the asymptotics of the Bessel and dilogarithm functions, in the conformal limit we obtain:
\begin{align}
\text{q}_1 \simeq \text{j}_{-1} &\simeq \frac{m^2}{2\pi} \sum_{n=1}^{\infty} (-1)^{n+1}\l\frac{1}{(nr_L)^2}+\frac{1}{(nr_R)^2}\r \nonumber \\ &= \frac{m^2 \eta(2)}{2\pi} \l \frac{1}{r_L^2}+\frac{1}{r_R^2}\r = \frac{\pi}{24} \l \frac{1}{\beta_L^2}+\frac{1}{\beta_R^2}\r, \label{ff_energy_density_asymptotics}\\
\text{q}_{-1} = \text{j}_{1} &\simeq \frac{m^2}{2\pi} \l\frac{\pi^2}{12r_L^2}-\frac{\pi^2}{12r_R^2}\r = \frac{\pi}{24} \l \frac{1}{\beta_L^2}-\frac{1}{\beta_R^2}\r \label{ff_energy_current_asymptotics},
\end{align} 
where 
\be
\eta(s) \equiv \sum_{n=1}^{\infty}\frac{(-1)^{n+1}}{n^s} = (1-2^{1-s})\zeta(s),
\ee 
is the Dirichlet $\eta$ function and we used $\eta(2)=\frac{\pi^2}{12}$. Since the free fermion central charge is $c=\frac{1}{2}$, the CFT results \eqref{CFT_energy_results} are recovered. 

We now turn to the higher-spin charges in the theory. The local conserved charges of the free fermion can take all integer values of the spin $s$. The charge eigenvalues with defined parity are
\be
h_s (\vartheta) = m^s \cosh (s\vartheta), \quad h_{-s} (\vartheta) = m^s \sinh (s\vartheta), \quad s \in \mathbb{N},
\ee
in agreement with \eqref{eigenvalues_even_odd_definition}. In terms of the integrals introduced in \eqref{ff_integrals_definition}, the average charge and current densities read:

\begin{align}
\text{q}_s &= \int \frac{\mathrm{d}\vartheta}{2\pi} n(\vartheta) e(\vartheta) h_s(\vartheta) = \frac{m^{s+1}}{2\pi} [\mathcal{I}_{1,s}^{++}(r_L)+\mathcal{I}_{1,s}^{++}(r_R)], \\
\text{q}_{-s} &= \int \frac{\mathrm{d}\vartheta}{2\pi} n(\vartheta) e(\vartheta) h_{-s}(\vartheta) = \frac{m^{s+1}}{2\pi} [\mathcal{I}_{1,s}^{+-}(r_L)-\mathcal{I}_{1,s}^{+-}(r_R)], \\
\text{j}_s &= \int \frac{\mathrm{d}\vartheta}{2\pi} n(\vartheta) p(\vartheta) h_s(\vartheta) = \frac{m^{s+1}}{2\pi} [\mathcal{I}_{1,s}^{-+}(r_L)-\mathcal{I}_{1,s}^{-+}(r_R)], \\
\text{j}_{-s} &= \int \frac{\mathrm{d}\vartheta}{2\pi} n(\vartheta) p(\vartheta) h_{-s}(\vartheta) = \frac{m^{s+1}}{2\pi} [\mathcal{I}_{1,s}^{--}(r_L)+\mathcal{I}_{1,s}^{--}(r_R)].
\end{align}
Only the integrals of the type $\mathcal{I}_{\alpha, \beta}^{++}$ and $\mathcal{I}_{\alpha, \beta}^{--}$ admit closed-form expressions as series of modified Bessel functions. These yield:
\begin{align}
\text{q}_s &= \frac{m^{s+1}}{2\pi} \sum_{n=1}^{\infty}(-1)^{n+1}\left[K_{s+1}(nr_L) - \frac{s}{nr_L}K_s(nr_L) + K_{s+1}(nr_R) - \frac{s}{nr_R}K_s(nr_R)\right], \\
\text{j}_{-s} &= \frac{m^{s+1}}{2\pi} \sum_{n=1}^{\infty}(-1)^{n+1}\left[\frac{s}{nr_L}K_s(nr_L) + \frac{s}{nr_R}K_s(nr_R)\right], \label{j_-s free fermion}
\end{align}
with asymptotics:
\be
\label{ff_higher_density_asymptotics}
\text{q}_s \simeq \text{j}_{-s} \simeq \frac{m^{s+1}}{4\pi} 2^s s! \sum_{n=1}^{\infty} \l\frac{1}{(n r_L)^{s+1}}+\frac{1}{(n r_R)^{s+1}}\r = \frac{2^s s! \eta(s+1)}{4\pi} \l\frac{1}{\beta_L^{s+1}} + \frac{1}{\beta_R^{s+1}}\r,
\ee
in the conformal limit $r_L$, $r_R \to 0$. We mention that the result \eqref{j_-s free fermion} was already derived in Appendix B of \cite{staircase}, with a slightly different notation. Although the averages $\text{q}_{-s}$, $\text{j}_s$ do not have finite-temperature closed-form expressions in terms of Bessel functions, in Appendix \ref{Appendix:Free fermion integrals and finite temperature corrections} we show that in the conformal limit the expected asymptotics are recovered:
\be
\label{ff_higher_current_asymptotics}
\text{q}_{-s} \simeq \text{j}_s \simeq \frac{2^s s! \eta(s+1)}{4\pi} \l\frac{1}{\beta_L^{s+1}} - \frac{1}{\beta_R^{s+1}}\r.
\ee
\subsection{Inhomogeneous GGE}
\label{InGGE}
By choosing the generalised thermodynamic potentials as $\beta_s = \beta^s$, the free fermion TBA equations for the two asymptotic reservoirs are
\be
\varepsilon_{R/L}(\vartheta) = \sum_s \beta_{R/L}^s h_s(\vartheta) = \sum_s r_{R/L}^s \cosh(s\vartheta),
\ee
and thus the GGE occupation function in the NESS is:
\be
n(\vartheta) = \frac{1}{1+ e^{\sum_s r^s_L \cosh{(s\vartheta)}}}\Theta(\vartheta)+  \frac{1}{1+ e^{\sum_s r_R^s \cosh{(s\vartheta)}}}\Theta(-\vartheta).
\ee
However, even for the free fermion the computation of expectation values using the full GGE occupation function is quite hard and does not lead to closed-form expressions. We therefore consider again the case of asymptotic reservoirs in spin-$s$ states, thus:
\be
\varepsilon_{s,R/L}(\vartheta) =  r_{R/L}^s \cosh(s \vartheta), \quad s \in \mathbb{N},
\ee
and 
\be
n_s(\vartheta) = \frac{1}{1+ e^{ r^s_L \cosh{(s\vartheta)}}}\Theta(\vartheta)+  \frac{1}{1+ e^{r_R^s \cosh{(s\vartheta)}}}\Theta(-\vartheta).
\ee
In this GGE we can compute the density and current averages of the charges with one-particle eigenvalue $h_{\pm k}(\vartheta)$, even when the spin $k \ne s$. Indeed, the spin-$k$ average charge density in a spin-$s$ state is expressed in terms of the integrals \eqref{ff_integrals_definition} by means of a simple change of variable $\vartheta \mapsto \frac{\vartheta}{s}$:
\begin{align}
\text{q}_k^{(s)} = \int \frac{\mathrm{d}\vartheta}{2\pi} n_s(\vartheta) e(\vartheta) h_k(\vartheta) &= \frac{m^{k+1}}{2\pi s} \int \frac{\mathrm{d}\vartheta}{2\pi}\, n_s\l\frac{\vartheta}{s}\r \cosh\l\frac{\vartheta}{s}\r \cosh\l\frac{k}{s}\vartheta\r \nonumber \\&= \frac{m^{k+1}}{2\pi s}\left[\mathcal{I}_{\frac{1}{s},\frac{k}{s}}^{++} (r^s_L)+\mathcal{I}_{\frac{1}{s},\frac{k}{s}}^{++} (r^s_R)\right].
\end{align}
Analogously,
\begin{align}
\text{q}_{-k}^{(s)} &=\frac{m^{k+1}}{2\pi s}\left[\mathcal{I}_{\frac{1}{s},\frac{k}{s}}^{+-} (r^s_L)-\mathcal{I}_{\frac{1}{s},\frac{k}{s}}^{+-} (r^s_R)\right], \\
\text{j}_k^{(s)} &=\frac{m^{k+1}}{2\pi s}\left[\mathcal{I}_{\frac{1}{s},\frac{k}{s}}^{-+} (r^s_L)-\mathcal{I}_{\frac{1}{s},\frac{k}{s}}^{-+} (r^s_R)\right], \\
\text{j}_{-k}^{(s)} &=\frac{m^{k+1}}{2\pi s}\left[\mathcal{I}_{\frac{1}{s},\frac{k}{s}}^{--} (r^s_L)+\mathcal{I}_{\frac{1}{s},\frac{k}{s}}^{--} (r^s_R)\right],
\end{align}
and in the conformal limit, the small-argument expansion \eqref{Bessel_small_argument_expansion} of the Bessel functions yields
\begin{align}
\label{q_k^s_ff}
\text{q}_k^{(s)} \simeq \text{j}_{-k}^{(s)} &\simeq \frac{2^{\frac{1+k}{s}}}{8\pi s} \Gamma \l\frac{1+k}{s}\r\eta\l\frac{1+k}{s}\r\l \frac{1}{\beta_L^{1+k}} + \frac{1}{\beta_R^{1+k}}\r, \\
\label{j_k^s_ff}
\text{q}_{-k}^{(s)} \simeq \text{j}_{k}^{(s)} &\simeq \frac{2^{\frac{1+k}{s}}}{8\pi s} \Gamma \l\frac{1+k}{s}\r\eta\l\frac{1+k}{s}\r\l \frac{1}{\beta_L^{1+k}} - \frac{1}{\beta_R^{1+k}}\r.
\end{align}
Interestingly, the spin $s$ of the GGE charge appears only in the coefficient: at the leading order, the temperature power law depends only on the spin $k$ of the charge which is averaged over the ensemble. 

In order to compare the free fermion results with the general expressions obtained in Section \ref{Section: main results}, we compute the coefficients $\mathcal{C}(s)$ and $\widetilde{\mathcal{C}}(s)$. The free fermion function $L(\vartheta)$ can be expanded in powers of $\exp(-r\cosh{\vartheta})$, hence:
\begin{align}
\int \mathrm{d}\vartheta \cosh (s\vartheta)L(\vartheta) &= \int_{-\infty}^{\infty} \mathrm{d}\vartheta \, \cosh{(s \vartheta)}\ln \l1+e^{-r \cosh \vartheta} \r \nonumber \\ &= \sum_{n=1}^\infty \frac{(-1)^{n+1}}{n}\int_{-\infty}^{\infty} \cosh{(s \vartheta)} e^{-n r \cosh \vartheta} = 2 \sum_{n=1}^\infty\frac{(-1)^{n+1}}{n} K_s(nr),
\end{align}
and the coefficient $\mathcal{C}(s)$ is:
\be
\label{C(s)_ff}
\mathcal{C}(s) = \lim_{r \to 0} \frac{6 r^s}{2^{s-1} \pi^2} \sum_{n=1}^\infty\frac{(-1)^{n+1}}{n} K_s(n r) = \frac{6}{\pi^2} \Gamma(s) \eta(s+1).
\ee
In a completely analogous way:
\begin{align}
\int \mathrm{d}\vartheta \cosh (\vartheta)L_s(\vartheta) &= \int_{-\infty}^{\infty} \mathrm{d}\vartheta \, \cosh{\vartheta}\ln \l1+e^{-r^s \cosh(s \vartheta)} \r \nonumber \\ &=  \frac{2}{s} \sum_{n=1}^\infty\frac{(-1)^{n+1}}{n} K_{\frac{1}{s}}(n r^s),
\end{align}
and 
\be
\widetilde{\mathcal{C}}(s) = \lim_{r \to 0} \frac{6 r}{s \pi^2} \sum_{n=1}^\infty\frac{(-1)^{n+1}}{n} K_{\frac{1}{s}}(n r^s) = 2^\frac{1}{s}\frac{3}{\pi^2} \Gamma\l 1+\frac{1}{s}\r \eta \l 1+\frac{1}{s}\r.
\ee
By inserting the expression for $\mathcal{C}(s)$ in equations \eqref{q_s NESS}-\eqref{j_s NESS}, one recovers the free fermion results \eqref{ff_higher_density_asymptotics}- \eqref{ff_higher_current_asymptotics}, and similarly the results \eqref{q_k^s_ff}-\eqref{j_k^s_ff} in the special case $k=s$ are recovered by inserting the free fermion expression for $\widetilde{\mathcal{C}}(s)$ in \eqref{q^s_s e j^s_s NESS}. As mentioned in the previous Section, $\mathcal{C}(s)$ is monotonically increasing to infinity as $s$ increases, while $\widetilde{\mathcal{C}}(s)$ decreases to the constant value $\frac{3 \ln 2}{\pi^2}$.

As a final remark, we show that in the case of a free fermion, it is possible to at least formally express $\mathcal{C}(s)$ as a sum of polylogarithms. Indeed, since in a free theory $e^\vartheta = \varepsilon^+(\vartheta) = (\varepsilon^{+})'(\vartheta)$, one can exactly evaluate the shifted integral in \eqref{C(s)_definition} by performing a change of variable:
\begin{align}
&\frac{6}{\pi^2} \int_{-x}^{+\infty}\mathrm{d}\vartheta\,e^{s\vartheta} L^+ (\vartheta) = \frac{6}{\pi^2} \int_{-x}^{+\infty}\mathrm{d}\vartheta\, (\varepsilon^+(\vartheta))^{s-1} \varepsilon^{\prime +}(\vartheta) L^+ (\vartheta) \nonumber \\= &\frac{6}{\pi^2} \int_{\varep_0}^{+\infty}\mathrm{d}\varep\, \varep^{s-1} \ln (1+e^{-\varep}) = - \frac{6}{\pi^2} \sum_{p=2}^{s+1}\frac{(s-1)!}{(s+1-p)!}\varep_0^{s+1-p}\text{Li}_p (-e^{-\varep_0}),
\end{align}
where $\varep_0 \equiv \varep(0)$ and the polylogarithm function is:
\be
\text{Li}_p (z) = \sum_{n=1}\frac{z^n}{n^p}.
\ee
The quantity $\mathcal{C}(s)$ is obtained by taking the $x \to \infty$ limit of the previous expression, and because at large temperatures the free fermion solution of the constant TBA equation \cite{klassen1990,Klassen1991} is  $\varep(0)=0$, the previous expression is only formal, as it reduces to \eqref{C(s)_ff}. Nonetheless, this calculation shows in which sense --at least in the case of a free theory-- the coefficient of the thermal averages for spin $s>1$ can be considered, in the CFT limit, a generalisation of the effective central charge, the latter being expressed as a sum of dilogarithms, with arguments given by (constant) TBA data.

\section{Numerical Results}
\label{Numerical results}
In this Section, we provide a numerical confirmation of our predictions for the average NESS conserved current densities in the conformal limit of some massive IQFTs. We do so by first solving the thermal TBA equations \eqref{TBA_eq_thermal} for the left and right reservoir at dimensionless scales $r_L$, $r_R$. The numerical solution of a single-particle TBA equation with driving term $w(\vartheta)$ is achieved as standard via successive iterations, starting with an initial function $\varep_0(\vartheta) = w(\vartheta)$ and then computing:
\be
\label{TBA_iteration}
\varep_i(\vartheta) = w(\vartheta) - \varphi * L_w[\varep_{i-1}](\vartheta), \quad i \in \mathbb{N},
\ee
until the process converges to the actual pseudoenergy $\varep(\vartheta) = \lim_{i\to \infty}\varep_i(\vartheta)$. The same process is applied to solve the dressing equation \eqref{dressing_definition}. Once the occupation functions of the two reservoirs are known, we obtain the NESS occupation function \eqref{occupation_function_NESS} by numerically solving the Riemann problem. The iteration process in this case is slightly different, as there are two coupled equations that must be considered simultaneously. We start by considering an initial zero $\vartheta_*^0$ of the effective velocity, through which we find the occupation function $n^0(\vartheta)$. The latter is used to compute the effective velocity, that has now a zero at $\vartheta = \vartheta_*^1$. Using this value we construct $n^1(\vartheta)$ and the process is repeated until simultaneous convergence of $\vartheta^i_*$ and $n^i(\vartheta)$ to $\vartheta_*$ and $n(\vartheta)$ respectively. This usually requires a very small number of iterations. 
The knowledge of the off-equilibrium occupation function is sufficient to compute the thermal currents $\text{j}_s$ for different values of $s$ and plot them at different temperatures. In performing the UV limit, we keep a fixed ratio of the temperatures, that is we set $r_L = \sigma r_R$, with $\sigma < 1$ constant (so that $T_L > T_R$). 

\begin{figure}
\centering
\begin{subfigure}[b]{\textwidth}
   \includegraphics[width=1\linewidth]{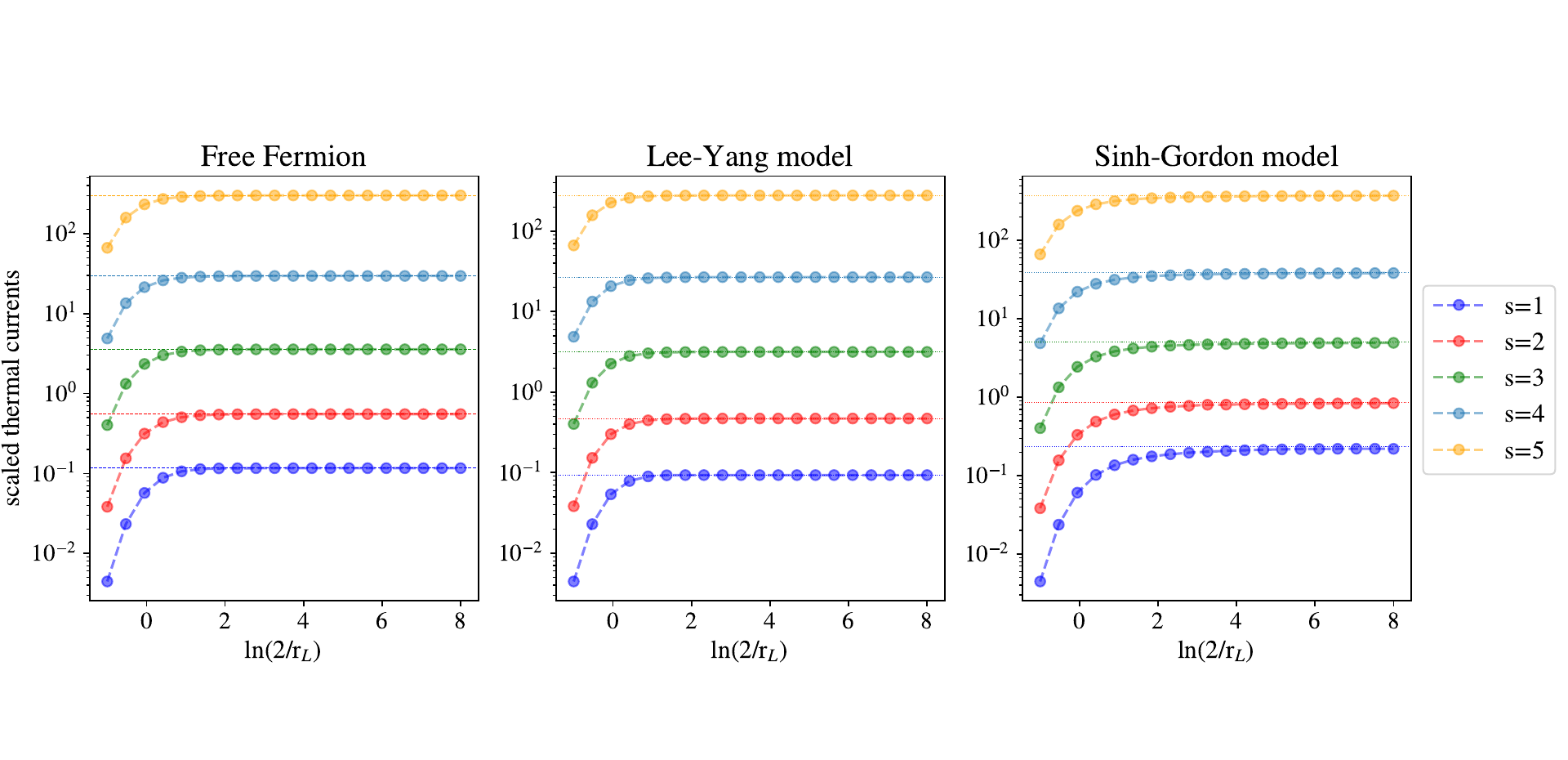}
   \caption{Scaled current averages $\beta_L^{s+1}\text{j}_s$. The dashed horizontal lines are at $\frac{s 2^{s-1}\pi}{12}\mathcal{C}(s)(1-\sigma^{s+1})$.}
   \label{fig:thermal currents} 
\end{subfigure}
\begin{subfigure}[b]{\textwidth}
   \includegraphics[width=1\linewidth]{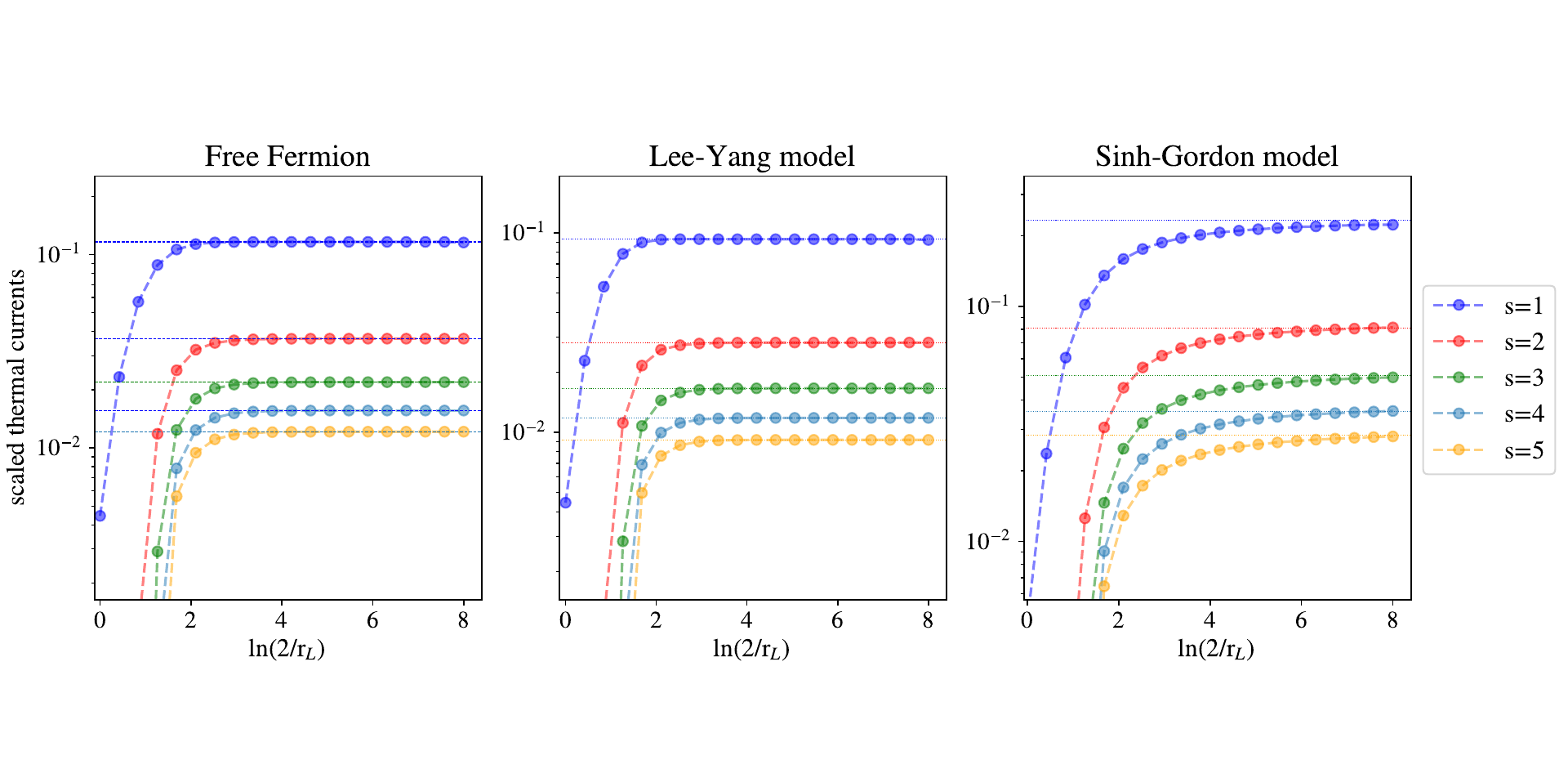}
   \caption{Scaled current averages $\beta_L^{s+1}\text{j}_s^{(s)}$. The dashed horizontal lines are at $\frac{\pi}{ 12 s }\widetilde{\mathcal{C}}(s)(1-\sigma^{s+1})$.}
   \label{fig:higher currents}
\end{subfigure}
\caption{Scaled (even) average currents in non-equilibrium thermal and spin-$s$ states for $s=1,2,3,4,5$ in the free fermion, sinh-Gordon and scaling Lee-Yang model. The ratio $\sigma = \frac{r_L}{r_R}$ is fixed at $\sigma=\frac{1}{3}$ and $x_L=\ln \frac{2}{r_L}$ is varied.}
\end{figure}

\begin{figure}
\centering
\begin{subfigure}[b]{\textwidth}
   \includegraphics[width=1\linewidth]{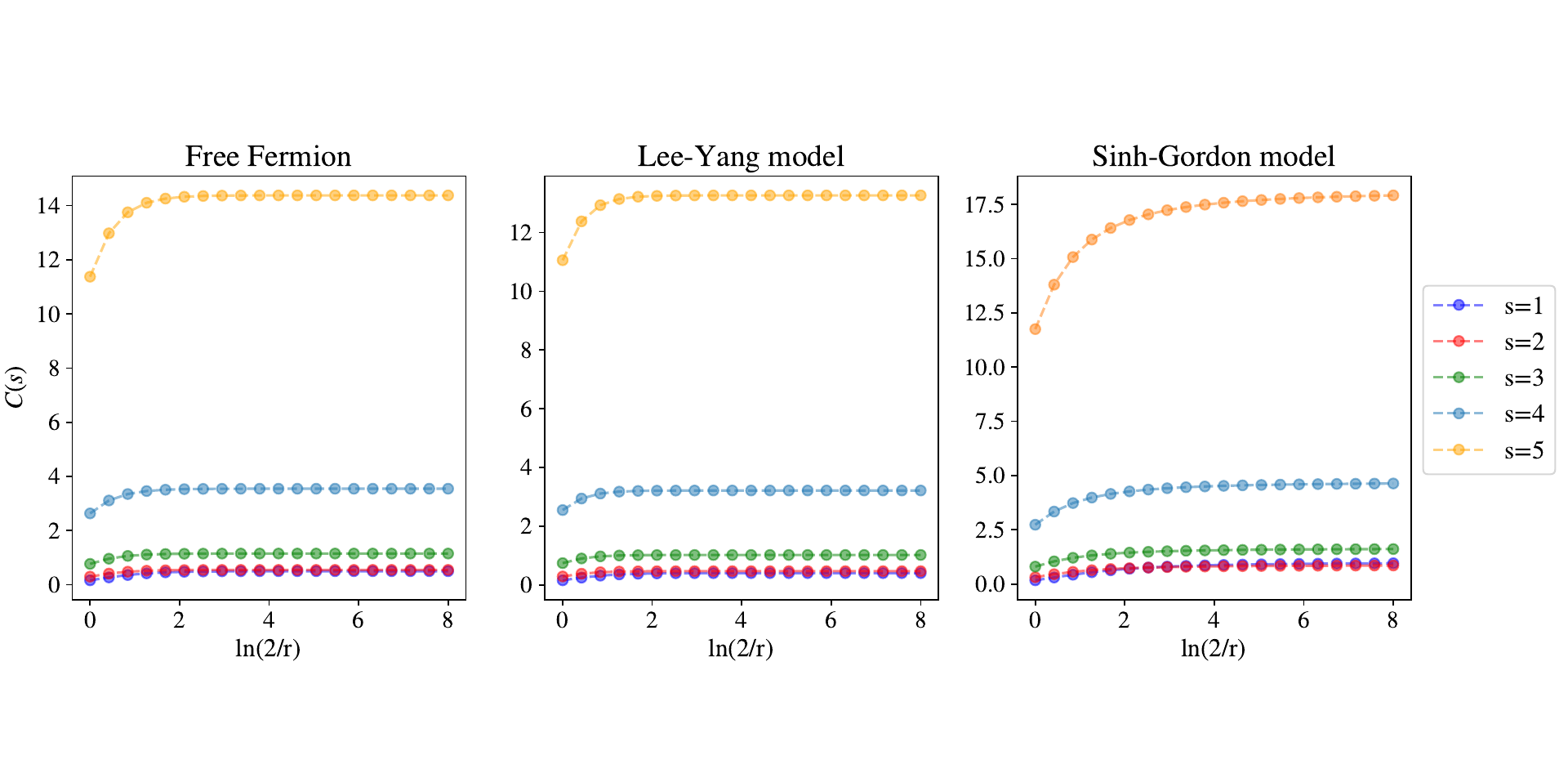}
   \caption{Coefficients  $\mathcal{C}(s)$.}
   \label{fig:coefficients C} 
\end{subfigure}
\begin{subfigure}[b]{1\textwidth}
   \includegraphics[width=1\linewidth]{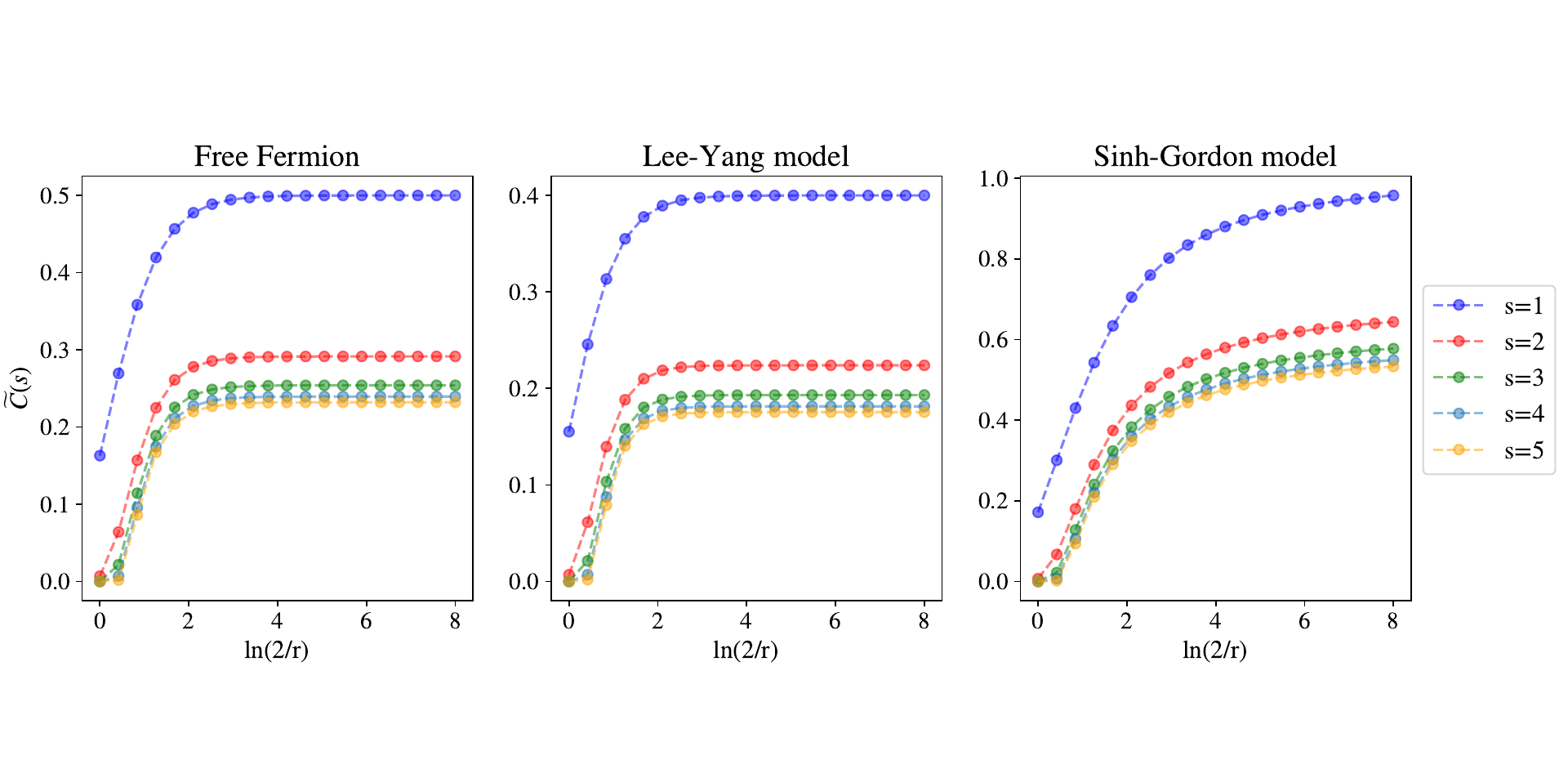}
   \caption{Coefficients $\widetilde{\mathcal{C}}(s)$.}
   \label{fig:coefficients C tilde}
\end{subfigure}
\caption{Coefficients $\mathcal{C}(s)$ and $\widetilde{\mathcal{C}}(s)$ for $s=1,2,3,4,5$ in the free fermion, sinh-Gordon and scaling Lee-Yang model as $x=\ln\frac{2}{r}$ varies. As expected, plateaux values are reached in the UV limit $x \gg 1$.}
\end{figure} 

We perform the numerical simulations on three single-particle theories: the free fermion, which we already considered in the previous Section, the sinh-Gordon model at the self-dual point and the scaling Lee-Yang model. The latter are interacting theories with scattering kernels given by:
\be   
\varphi_\text{ShG}(\vartheta) = \frac{2}{\cosh(\vartheta)}, \quad \varphi_\text{LYM}(\vartheta) = -\frac{4\sqrt{3} \cosh(\vartheta)}{1+2\cosh(2\vartheta)}.
\ee
The TBA of these models was first studied in \cite{zamolodchikov2006thermodynamic} and \cite{Zamolodchikov1991tba} respectively. The conformal limit of the sinh-Gordon model is a free boson with $c=1$, while the Lee-Yang scaling model flows to the non-unitary minimal model $\mathcal{M}_{2,5}$, with effective central charge $c_\text{eff}=\frac{2}{5}$. In Figure \ref{fig:thermal currents} we plot the scaled thermal currents $\beta_L^{s+1}\text{j}_s$, with $\text{j}_s$ given by \eqref{average thermal current split}. As expected, in the conformal limit, with $\frac{\beta_L}{\beta_R}$ fixed, the scaled currents are independent of $\beta_L$ and reach the $s$-dependent constant predicted by \eqref{j_s NESS}. In figure \ref{fig:higher currents} we plot the quantities $\beta_L^{s+1}\text{j}^{(s)}_s$, where now $\text{j}^{(s)}_s$ are the higher-spin current densities in spin-$s$ states. In this case, we observe that the scaled currents asymptotically reach the values predicted by \eqref{q^s_s e j^s_s NESS}. Interestingly, deviations from the constant values are very small also in the sinh-Gordon model, although famously in this theory the TBA functions $L(\vartheta)$ and $n(\vartheta)$ do not display a plateau-like structure as the one depicted in Fig.~\ref{fig:equilibrium_occupation_function}, but are bell-shaped functions of $\vartheta$ \cite{zamolodchikov2006thermodynamic}. This property makes the numerical convergence of the integrals in this theory significantly worse. It is important to stress that in the Lee-Yang scaling model, the quantities 
$\text{j}_s$ and $\text{j}_s^{(s)}$ for $s = 2,3,4$ do not correspond to conserved physical currents, as the first higher-spin local conserved charge of the model is at $s=5$ \cite{cardy1989s}. Similarly, in the sinh-Gordon model there exist only local conserved quantities for odd spins. Nevertheless, the integrals can formally be computed and allow us to confirm the predicted scaling laws, which are the main result of this paper.

In Figures \ref{fig:coefficients C} and \ref{fig:coefficients C tilde} we plot the coefficients $\mathcal{C}(s)$ and $\widetilde{\mathcal{C}}(s)$, as defined by equations \eqref{C(s)_definition} and \eqref{Tilde_C(s)_definition}. The plateaux values at large $x$ are the ones used to compute the asymptotic values in Fig.~\ref{fig:thermal currents} and \ref{fig:higher currents}, and in the free fermion case they coincide with the ones computed in the previous Section. The $s=1$ values are precisely the central charges (or effective central charges) of the UV fixed points, namely $\mathcal{C}(1)=\widetilde{\mathcal{C}}(1) = \frac{1}{2}, 1, \frac{2}{5}$ for the free fermion, sinh-Gordon and Lee-Yang model, respectively. As expected, the plots show the exponential increase of $\mathcal{C}(s)$ with larger values of $s$ and the corresponding decrease of $\widetilde{\mathcal{C}}(s)$.

\section{Conclusions and Outlook}
\label{Conclusions and Outlook}
In this paper, we obtained the expressions of average currents and densities of higher-spin local conserved charges in out-of-equilibrium (1+1)-dimensional CFTs. The non-equilibrium setting we considered is that of a partitioning protocol, in which the two halves of the system at $x>0$ and $x<0$ are initially prepared either in different thermal states or in different generalised Gibbs states. The averages we computed are those arising at large times in the NESS at $x=0$.

Our results were obtained using the quasi-particle description provided by the TBA in conjunction with the hydrodynamic principles underlying GHD. We computed expectation values of  higher-spin observables in massive IQFTs and derived the corresponding conformal predictions in the UV (or zero mass) limit. In doing so, we reproduced first of all the well-known CFT results for the scaling of the energy current and charge densities in the NESS: if $T_L$, $T_R$ are the temperatures of the two thermal reservoirs, then the energy density is proportional to $T_L^2 + T_R^2$ and the energy current is proportional to $T_L^2 - T_R^2$, with a coefficient proportional to the (effective) central charge of the CFT. 

Our first original result was to show that these scaling laws naturally generalise to other (local) conserved charges besides the energy and momentum: for a charge of spin $s$ and thermal reservoirs, the NESS average charge density and current are proportional to $T_L^{s+1} \pm T_R^{s+1}$, with a coefficient $\mathcal{C}(s)$, which we can think of as the UV limit of a generalised scaling function of the massive IQFT. 

In addition to this result, we obtained the same  CFT averages for asymptotic reservoirs which are characterised by a single non-vanishing potential coupled to a higher-spin charge of spin $s$ in the state. We called this particular type of GGEs, spin-$s$ states. In such a state, the generalised temperature in the TBA equations is chosen as $\beta^s$, with $\beta$ an inverse temperature. With this choice, we showed that in the partitioning protocol the scaling $T_L^{s+1} \pm T_R^{s+1}$ for the average density and current associated to a conserved quantity of spin $s$ is preserved, with a different proportionality coefficient  $\widetilde{\mathcal{C}}(s)$. 
Furthermore, we have shown that the exact dependence of the CFT expectation values on the effective central charge is restored when the dynamics of the massive IQFT is ruled by a higher-spin charge $Q_s$, rather than the Hamiltonian, and thus the continuity equations are defined according to a generalised time $t_s$ associated to $Q_s$. 

For a relativistic massive free fermion we obtained exact formulae for the average densities and currents for generic values of the spin, mass and temperature. In the CFT limit, these results agree with the general scaling laws obtained for interacting theories. Using the TBA equation of a massive free fermion we obtained an alternative representation of the coefficient $\mathcal{C}(s)$, which sheds some light on its relation to dilogarithms as are typically found for TBA scaling functions. 

Finally, we have numerically verified our results by solving the TBA equations and the partitioning protocol for the sinh-Gordon and the Lee-Yang models. In the UV limit, the numerical results are in excellent agreement with our predictions for the scaling laws in CFTs.

\medskip 
Several research directions can be pursued starting from the results presented in this work. First, it would be interesting to test the robustness of the temperature power laws that we have identified for systems which are initially prepared in GGEs with a single non-vanishing generalised potential. Our intuition and current results both suggest that the scaling is dictated only by the spin of the charge which is averaged, and does not depend on the state, even if the overall coefficients do.
To check this prediction one would need to numerically solve the TBA equation with a driving term containing several charge eigenvalues, a computationally demanding task. 

Second, aside from the temperature dependence of the expectation values, further information on the coefficients $\mathcal{C}(s)$ and $\widetilde{\mathcal{C}}(s)$ could be obtained by computing the averages directly at the CFT point by using finite-size techniques. For instance, higher-spin conserved densities are obtained by considering the descendants of the stress-energy tensor, which yield non-vanishing expectation values when mapped to a strip \cite{blote1986conformal, affleck1988universal}. The algebra of these operators on a finite-size geometry could reveal the exact dependence of the finite-temperature averages on the central charge and on the spin, at least for some families of conserved charges.

Third, in this paper, we looked only at the expectation values of the observable in the state at $\xi=0$, with a focus on the high temperature regime. However, it is known that in the partitioning protocol, when the temperatures of the two half-lines are small but different, there is a light-cone broadening effect  for systems admitting an effective low-energy description by means of a non-linear Luttinger liquid \cite{Lorenzo1}, with profiles of the currents showing smooth peaks in $\xi$ around the edges of the light-cone. This effect was also observed in the gapless regime of the XXZ chain \cite{Lorenzo2}. At least in the case of the relativistic massive free fermion, we have access to closed analytic expressions for the charge and current profiles at every value of the mass, temperature and in a partitioning protocol -although we have not presented this result here- also of the ray $\xi$. Therefore, beside a numerical study which could be performed for integrable theories, the free-fermion model can be used to analytically check whether these low-temperature phenomena appear also in relativistic theories and to characterise the transition between the gapped and the gapless phase.

Fourth, it would be interesting to characterise the full-counting statistics associated to transport phenomena in the partitioning protocol. The generating function of the cumulants for the energy transfer at large times in CFT was computed in \cite{bernard2012energy} and later in \cite{Bernard_2014,bernard2016conformal}. The computation of the cumulants reduce to that of many-point correlation functions of the current densities, a problem which was solved using GHD techniques within the framework of ballistic fluctuation theory in \cite{myers2020transport}. Following the protocol developed therein, a natural extension of this work would be to compute the large-deviation functions of higher-spin currents in massive theories and then take the UV limit in order to obtain the full counting statistic of those currents in CFT, generalising the results of \cite{bernard2012energy}.

Finally, we mention a direction of research we intend to pursue in the very near future: this is the study of charge densities and currents in gapped theories perturbed by irrelevant operators. There is a vast literature regarding the effects induced by the irrelevant $\TTb$-deformation of IQFTs (see e.g. \cite{Zamolodchikov:2004ce, Smirnov:2016lqw,Cavaglia:2016oda}), and the behaviour of $\TTb$-deformed CFTs in off-equilibrium settings was recently studied in \cite{Medenjak:2020bpe,Medenjak:2020ppv}. The formalism and the results derived in the present paper can be generalised to $\TTb$-perturbed IQFTs. This study is the subject of our upcoming 
work \cite{OurUnpublished}.

\medskip

 \noindent {\bf Acknowledgements:} 
 The authors thank Benjamin Doyon, Gerard Watts and Davide Fioravanti for useful discussions. They are especially grateful to Benjamin Doyon for clarifying some doubts about the physical implications of the results presented in this paper.  Michele Mazzoni is grateful for funding under the EPSRC Mathematical Sciences Doctoral Training Partnership EP/W524104/1.  Riccardo Travaglino thanks the Department of Mathematics at City, University of London for hospitality during a three month visit in 2023, when part of this project was completed. Michele and Riccardo thank the organisers of the SFT 2024 Lectures on Statistical Field Theories (February 2024) at GGI Florence (Italy), where final work on this draft was carried out. Olalla A. Castro-Alvaredo thanks EPSRC for financial support under Small Grant EP/W007045/1.

\appendix

\section{Useful Bounds on the Effective Velocity}
\label{appendix:Useful bounds on the effective velocity}
It is not difficult to show that the effective velocity of an interacting theory does not deviate too much from the $\tanh \vartheta$ function of the free case. This was already observed numerically for the sinh-Gordon model in \cite{castro2016emergent}. In fact, it is known that the effective velocity satisfies the self-consistent equation \cite{castro2016emergent}:
\begin{equation}
    v^{\text{eff}}(\vartheta) = \tanh\vartheta + \int \mathrm{d}\alpha \,\frac{\varphi(\vartheta-\alpha)\rho_p(\alpha)}{\cosh(\vartheta)}( v^{\text{eff}}(\alpha) -  v^{\text{eff}}(\vartheta)),
\end{equation}
where $\rho_p(\vartheta)$ is the TBA density of occupied states per unit length. This immediately allows us to write:
\begin{eqnarray}
    v^{\text{eff}}(\vartheta) &=& \frac{\tanh\vartheta + \int \mathrm{d}\alpha\, \frac{\varphi(\vartheta-\alpha)\rho_p(\alpha)}{\cosh(\vartheta)}v^{\text{eff}}(\alpha)}{1+\int \mathrm{d}\alpha\,\frac{\varphi(\vartheta-\alpha)\rho_p(\alpha)}{\cosh(\vartheta)}} \nonumber\\ 
     &=& \frac{\sinh\vartheta + \int \mathrm{d}\alpha\, \varphi(\vartheta-\alpha)\rho_p(\alpha)v^{\text{eff}}(\alpha)}{\cosh\vartheta+\int \mathrm{d}\alpha\, \varphi(\vartheta-\alpha)\rho_p(\alpha)}.
\end{eqnarray}
Since we expect the theory to remain local and causal in the presence of interactions (a claim which is violated for example by theories in which the $\TTb$ deformation is present \cite{Medenjak:2020ppv,OurUnpublished}), the effective velocity will satisfy the bound $-1 \leq v^{\rm eff}(\alpha) \leq 1$. Substituting this in the above equation, we get:
\begin{equation}
    \frac{\sinh\vartheta - A(\vartheta)}{\cosh\vartheta+A(\vartheta)} \leq v^{\rm eff}(\vartheta) \leq \frac{\sinh\vartheta + A(\vartheta)}{\cosh\vartheta+A(\vartheta)}
\end{equation}
Where $A(\vartheta) = \int \varphi(\vartheta - \alpha) \rho_p(\alpha)$. In general, the kernels of realistic scattering theories are functions which go to zero exponentially fast at $\vartheta \to \pm \infty$ and have one or more maxima close to $\vartheta\approx 0$. Denoting by $\varphi^*$ the global maximum, we have
\begin{equation}
    A(\vartheta) \leq \varphi^* \int \mathrm{d}\alpha\, \rho_p(\alpha) = \varphi^* N_p,
\end{equation}
where $N_p$  is the number of quasi-particles excitations per unit length, which (at equilibrium) is fully determined by TBA data. This implies that the effective velocity of a generic theory, at any value of the scale $r$, deviates from the hyperbolic tangent characteristic of the free theory only in a small region around the origin. For instance, we see that the value of $\vartheta_*$ is constrained to lie in the segment:
\begin{equation}
    -\sinh^{-1}(\varphi^* N_p) \leq \vartheta_* \leq \sinh^{-1}(\varphi^* N_p).
\end{equation} 
This bound could be further refined by noting that, generally, $\rho_p(\vartheta)$ has two well defined peaks at $\vartheta_{\pm} \approx \log(2/r)$, while the kernel is peaked around the origin.

\section{Spin-Dependent Scaling Function}
\label{appendix:Spin-dependent scaling function}
In this appendix we prove that the spin-dependent scaling function \eqref{spin-s scaling function def} reproduces the effective central charge of the underlying CFT in the UV limit, i.e. we prove equation \eqref{spin-s scaling function limit}. Our proof follows the original argument by Zamolodchikov \cite{Zamolodchikov1991tba} for thermal TBA. In the simplest case of a theory with a single-particle spectrum, Zamolodchikov's showed how equation \eqref{scaling function limit Zamolodchikov} follows from the relation:
\beq
c_\text{eff}= \frac{6}{\pi^2}\mathcal{L}(n(0)),
\eeq
where $n(0)$ is the plateau value of $n(\vartheta)$, which can be obtained from the solution of the constant TBA equation:
\be
\label{constant TBA} \varep(0) = N \ln \l 1+e^{-\varep(0)} \r,\quad N = -\int_{-\infty}^\infty \frac{\mathrm{d}\vartheta}{2\pi}\,\varphi(\vartheta),
\ee
and $\mathcal{L}(x)$ is Roger's dilogarithm, which for $0<x<1$ has the integral representation:
\beq
\label{Roger dil}
\mathcal{L}(x)=\frac{1}{2}\int_x^0\mathrm{d}y\,\left[\frac{\ln{(1-y)}}{y}+ \frac{\ln{y}}{1-y}\right]\,.
\eeq
Let us consider the TBA equation for a single-particle theory in a spin-$s$ state. In the large $x$ limit, using the (left-) shifted equations \eqref{shifted_spin_s_TBA_eq} and \eqref{shifted_spin_s_TBA_eq_derivative} we can write
\beq
\label{e s theta normal}
e^{s\vartheta} \simeq \frac{1}{2^{s-1}}(\varep_s^+(\vartheta) + (\varphi*L_s^+)(\vartheta)),
\eeq
and 
\beq
\label{e s theta derivative}
e^{s\vartheta} \simeq \frac{1}{s 2^{s-1}}((\varep^+_s)'(\vartheta) + (\varphi*(L^+_s)')(\vartheta)).
\eeq
Using the second of these equations, the scaling function \eqref{spin-s scaling function def} can be cast as:
\begin{align}
\frac{3 s r^s}{\pi^2}\int_{-\infty}^{+\infty}\mathrm{d}\vartheta\,L_s(\vartheta)\cosh{(s\vartheta)}  &\simeq \frac{6 s 2^{s-1}}{\pi^2}\int_{-x}^{+\infty}\mathrm{d}\vartheta\,L_s^+(\vartheta)e^{s\vartheta} \nonumber \\  &\simeq \frac{6}{\pi^2}\int_{-x}^{+\infty}\mathrm{d}\vartheta\,L_s^+(\vartheta)(\varep^+_s)'(\vartheta) + \frac{6}{\pi^2}\int_{-x}^{+\infty}\mathrm{d}\vartheta\,L_s^+(\vartheta)(\varphi*(L^+_s)')(\vartheta)\,.
\end{align}
Changing variables from $\vartheta$ to $\varep_s^+$, the first integral in the second line becomes:
\beq
\frac{6}{\pi^2}\int_{\varep_s^+(-x)}^{\varep_s^+(\infty)}\mathrm{d}t\,\ln{(1+ e^{-t})},
\eeq
while the second can be rewritten as follows:
\begin{align}
\label{appendix_B_integral_splitting}
\frac{6}{\pi^2}\int_{-x}^{+\infty}\mathrm{d}\vartheta\,L_s^+(\vartheta)(\varphi*(L^+_s)')(\vartheta)&\simeq \frac{6}{\pi^2}\int_{-x}^{+\infty}\mathrm{d}\vartheta\,(L^+_s)'(\vartheta)(\varphi*L^+_s)(\vartheta) \nonumber \\
&\simeq 2^{s-1}\frac{6}{\pi^2}\int_{-x}^{+\infty}\mathrm{d}\vartheta\,(L^+_s)'(\vartheta)e^{s\vartheta} - \frac{6}{\pi^2}\int_{-x}^{+\infty}\mathrm{d}\vartheta\,(L^+_s)'(\vartheta)\varep_s^+(\vartheta),
\end{align}
where the first equality holds in the limit $x\to +\infty$ thanks to the symmetry of the scattering kernel and the second follows from \eqref{e s theta normal}. The first integral in the second line of \eqref{appendix_B_integral_splitting} can be done by parts, yielding:
\beq
2^{s-1}\frac{6}{\pi^2}\int_{-x}^{+\infty}\mathrm{d}\vartheta\,(L^+_s)'(\vartheta)e^{s\vartheta} \simeq -\frac{6 s 2^{s-1}}{\pi^2}\int_{-x}^{+\infty}\mathrm{d}\vartheta\,L^+_s(\vartheta)e^{s\vartheta},
\eeq
whereas a change of variable in the second integral gives:
\beq
-\frac{6}{\pi^2}\int_{-x}^{+\infty}\mathrm{d}\vartheta\,(L^+_s)'(\vartheta)\varep_s^+(\vartheta) =  \frac{6}{\pi^2}\int_{\varep^+_s(-x)}^{\varep_s^+(\infty)}\mathrm{d}t\, \frac{t}{1+e^t}.
\eeq
Putting all the pieces together and noting that $\varep_s^+(-x)=\varep_s(0)$, $\varep_s^+(\infty)=\infty$, we have:
\begin{align}
&\frac{6 s 2^{s-1}}{\pi^2}\int_{-x}^{+\infty}\mathrm{d}\vartheta\,L_s^+(\vartheta)e^{s\vartheta} \nonumber \\
\simeq & \frac{6}{\pi^2}\int_{\varep_s(0)}^{\infty}\mathrm{d}t\,\ln{(1+ e^{-t})} - \frac{6 s 2^{s-1}}{\pi^2}\int_{-x}^{+\infty}\mathrm{d}\vartheta\,L_s^+(\vartheta)e^{s\vartheta} + \frac{6}{\pi^2}\int_{\varep_s(0)}^{\infty}\mathrm{d}t\, \frac{t}{1+e^t}, 
\end{align}
which implies:
\begin{align}
\label{modified scaling function final}
\lim_{r \to 0} c_s(r) &=\lim_{x\to \infty} \frac{6 s 2^{s-1}}{\pi^2}\int_{-x}^{\infty}\mathrm{d}\vartheta\, L_s^+(\vartheta) e^{s\vartheta} \nonumber \\&= \frac{6}{\pi^2}\left[\frac{1}{2}\int_{\varep_s(0)}^{\infty}\mathrm{d}t\left(\ln{(1+e^{-t}) + \frac{t}{1+e^t}}\right)\right] = \frac{6}{\pi^2}\mathcal{L}(n_s(0)).
\end{align}
The same expression is obtained starting from the right-shifted spin-$s$ TBA equation. The term in the right-hand side of the last equality is precisely $c_\text{eff}$, since $n_s(0)=n(0)$. Indeed, the TBA equations \eqref{TBA_eq_thermal} and \eqref{spin-s_TBA_eq} have the same constant form, because for every $s > 0$ the term $r^s\cosh{(s\vartheta)}$ is exponentially suppressed in the region $-x \ll \vartheta \ll x$ as $x\to \infty$. This completes the proof.

\section{Many-Particle Theories}
The TBA formulation extends naturally and without relevant modifications to many-particle theories in which the scattering is diagonal, namely the $S$-matrix only has elements of the form $S_{ab}^{ab}(\vartheta_1,\vartheta_2) = S_{ab}(\vartheta_1-\vartheta_2)$. In this situation the spectrum is characterised by a set of particles labeled by $a=1,...,N$, having masses $m_a$ conventionally ordered so that $m_1$ is the smallest. There is a TBA equation for each particle, such that \eqref{GGE_TBA_general_definition} is modified to:

\begin{equation}
\label{eq:manyparticles_tba}
      \epsilon_a(\vartheta) = w_a(\vartheta) - \sum_b (\varphi_{ab}*L_b)(\vartheta), \quad w_a(\vartheta) = \sum_s \beta_s h_{a,s}(\vartheta), 
    \end{equation}
where the definition of the kernels is analogous to the single particle case: $\varphi_{ab}(\vartheta) = -i \frac{\partial}{\partial \vartheta} \ln{S_{ab}}(\vartheta)$. The presence of interactions leads to coupling of the different equations, introducing further nonlinearities which make the solution significantly more difficult than in the one-particle case. Of particular interest are theories where the structure of the TBA equations can be encoded into the Dynkin diagram of a semi-simple Lie algebra of the ADE type \cite{zamolodchikov1991ADE,DynkinTBA}:
\begin{equation}
    \epsilon_a(\vartheta) =  w_a(\vartheta) - \varphi * \sum_b G_{ab} (w_b - \epsilon_b - L_b)(\vartheta).
    \label{eq:universaldiagonal}
\end{equation}
In the equation above, $G$ is the adjacency matrix of the Dynkin diagram of some semi-simple Lie algebra, and we introduced the universal kernel $\varphi= \frac{g}{2 \cosh{\frac{g \vartheta}{2}}}$, with $g$ the dual Coxeter number of the algebra. In this situation each quasi-particle can be associated to a node in the diagram and the interactions between different nodes correspond to non-zero entries of $G$.

We now prove that the results obtained in Section \ref{Section: main results} naturally generalise to the diagonal many-particle case. Let us consider the shifted versions of equation \eqref{eq:manyparticles_tba} for a spin-$s$ state, $w_a(\vartheta)=m^s_a \beta^s \cosh(s\vartheta)$:
\begin{eqnarray}
    \varepsilon_{a}^{\pm}(\vartheta) &=& \hat{m}_a^s  2^{s-1} e^{\pm s}-\sum_b \varphi_{ab}*L_b^{\pm}(\vartheta), \\
    (\varepsilon_{a}^{\pm})'(\vartheta) &=& \pm s \hat{m}_a^s 2^{s-1} e^{\pm s\vartheta}+\sum_b \varphi_{ab}*(n_b^{\pm}(\varepsilon_{b}^{\pm})')(\vartheta),
\end{eqnarray}
Where for all $a$, $\varep_{a}^\pm(\vartheta)= \varep_a(\vartheta \pm x)$, $x= \ln \l\frac{2}{r}\r$ with $r\equiv m_1 \beta$ and $\hat{m}_a=m_a/m_1$. The dressing is expressed as:
\begin{equation}
\label{multi-particle dressing def}
    h_{a,s}^{\text{dr}}(\vartheta)=h_{a,s}(\vartheta) +\sum_b \varphi_{ab}* (n_b h_b^{\text{dr}})(\vartheta).
\end{equation}
Hence, as $h_{a,s}(\vartheta)=m_a^s\cosh(s\vartheta)$, the shifted dressing equation is also immediately generalised:
\be
    (h_{a,s}^{\pm})^\text{\text{dr}}= [\hat{m}_a^s\beta^{-s}2^{s-1}e^{\pm s\vartheta}]^{\text{dr}} = 2^{s-1}\beta^{-s}[\hat{m}_a^s e^{\pm s\vartheta}]^\text{\text{dr}}.
\ee
Note that the operation \eqref{multi-particle dressing def} is not separately linear in each component, hence for example in general $(m_a \cosh\vartheta)^{\text{dr}} \ne m_a (\cosh\vartheta)^{\text{dr}}$. However, the dressing is linear under multiplication of all the functions $h_a(\vartheta)$ by a constant, i.e. for instance $(A \cosh\vartheta)^{\text{dr}} = A (\cosh\vartheta)^{\text{dr}}$. The average charge densities are expressed as:
\begin{equation}
    \texttt{q}_s = \sum_a \int \frac{\mathrm{d}p_a}{2\pi} n_a(\vartheta) h_{a,s}^{\text{dr}}(\vartheta) = \sum_a m_a \int \frac{\mathrm{d}\vartheta}{2\pi} \cosh(\vartheta) n_a(\vartheta) h_{a,s}^{\text{dr}}(\vartheta).
\end{equation}
To study the partitioning protocol we introduce the left and right parts and shift the integrals as done previously:
\begin{eqnarray*}
    \texttt{q}_s &=& \sum_a \left[\int_{-\infty}^{\vartheta^*} \frac{\mathrm{d}\vartheta}{2\pi} m_a \cosh(\vartheta) n_{a,R}(\vartheta) h_{a,s}^{\text{dr}}(\vartheta) + \int_{\vartheta^*}^{\infty} \frac{\mathrm{d}\vartheta}{2\pi} m_a \cosh(\vartheta) n_{a,L}(\vartheta) h_{a,s}^{\text{dr}}(\vartheta)    \right]\\
   &=&  \sum_a \left[\int_{-\infty}^{\vartheta^*+x_R} \frac{\mathrm{d}\vartheta}{2\pi} e_a^-(\vartheta) n_{a,R}^{-}(\vartheta) h_{a,s}^{\text{dr},-}(\vartheta) + \int_{\vartheta^*-x_L}^{\infty} \frac{\mathrm{d}\vartheta}{2\pi} e_a^+(\vartheta) n_{a,L}^+(\vartheta) h_{a,s}^{\text{dr},+}(\vartheta)    \right] \\
    &=& \sum_a \left[\frac{2^{s-1}}{2\pi \beta_R^{s+1}}\int_{-\infty}^{\vartheta^*+x_R} \mathrm{d}\vartheta\, \hat{m}_a e^{-\vartheta} n_{R,a}^-(\vartheta) [\hat{m_a}^{s}e^{- s\vartheta}]^{\text{dr}} + \frac{2^{s-1}}{2\pi \beta_L^{s+1}}\int_{\vartheta^*-x_L}^{\infty} \mathrm{d}\vartheta \,\hat{m}_a e^{\vartheta} n_{L,a}^+(\vartheta) [\hat{m}_a^{s}e^{ s\vartheta}]^{\text{dr}}\right]\,.
\end{eqnarray*}
By making use of the same considerations already employed in the single particle case, we can invert the dressing and use the fact that $(\varepsilon_a^{\pm})' \propto (h^{\pm}_{a})^{dr} $ to recast the above as:
\begin{equation}
    \frac{s 2^{s-1}\pi}{12}\mathcal{C}^{(N)}(s)\l\frac{1}{\beta_L^{s+1}}+\frac{1}{\beta_R^{s+1}}\r,
\end{equation}
in the large $x$ limit. The quantity:
\be
\mathcal{C}^{(N)}(s)= \lim_{x\to \infty} \frac{6}{\pi^2} \sum_{a=1}^N \hat{m}_a \int_{-x}^{+\infty}\mathrm{d}\vartheta\,e^{s\vartheta} L_a^+ (\vartheta),
\ee
is the natural generalisation of the usual many-particle expression for the central charge (see for example \cite{Klassen1991}). A similar procedure can also be performed for the currents, thus confirming that the results of this work extend to the many-particle case. 

\label{Appendix:Many-particle theories}
\section{Free Fermion Integrals}
\label{Appendix:Free fermion integrals and finite temperature corrections}
We start by considering the integrals $\mathcal{I}^{++}_{\alpha \beta}(z)$ and $\mathcal{I}^{--}_{\alpha \beta}(z)$ defined in \eqref{ff_integrals_definition}. From the expansion (valid for any $z > 0$):
\be
\label{geometric series}
\frac{1}{1+e^{z \cosh \vartheta}} = \sum_{n=1}^{\infty} (-1)^{n+1}e^{-n z \cosh \vartheta},
\ee
and the identities:
\begin{align}
 \sinh{(\alpha \vartheta)} \sinh{(\beta \vartheta)} = \frac{1}{2}[\cosh{(\alpha \vartheta + \beta \vartheta)} - \cosh{(\alpha \vartheta - \beta \vartheta)}], \\
 \cosh{(\alpha \vartheta)} \cosh{(\beta \vartheta)} = \frac{1}{2}[\cosh{(\alpha \vartheta + \beta \vartheta)} + \cosh{(\alpha \vartheta - \beta \vartheta)}], \nonumber 
\end{align}
one gets
\begin{align}
\mathcal{I}^{++}_{\alpha \beta}(z) &= \frac{1}{2} \sum_{n=1}^\infty (-1)^{n+1} \int_0^\infty \mathrm{d}\vartheta \, [\cosh (\alpha \vartheta + \beta \vartheta) + \cosh (\alpha \vartheta - \beta \vartheta)] e^{-n z \cosh \vartheta} , \nonumber \\
\mathcal{I}^{--}_{\alpha \beta}(z) &= \frac{1}{2} \sum_{n=1}^\infty (-1)^{n+1} \int_0^\infty \mathrm{d}\vartheta \, [\cosh (\alpha \vartheta + \beta \vartheta) - \cosh (\alpha \vartheta - \beta \vartheta)] e^{-n z \cosh \vartheta},
\end{align}
and thus \eqref{modified_bessel_definition} yields:
\be
\mathcal{I}^{\pm \pm}_{\alpha \beta}(z) = \frac{1}{2} \sum_{n=1}^\infty (-1)^{n+1} \left[ K_{\alpha + \beta}(n z) \pm K_{\alpha - \beta} (n z) \right] .
\ee
From the small $z$ expansions:
\be
\label{Bessel_small_argument_expansion}
K_s(z) = \frac{2^{s-1}\Gamma(s)}{z^s} + \mathcal{O}(z^{-s+2}) \quad \text{for} \, s > 0, \quad K_0 = -\log\l\frac{z}{2}\r - \gamma_E + \mathcal{O}(z^2),
\ee
we obtain the results \eqref{ff_energy_density_asymptotics} and \eqref{ff_higher_density_asymptotics} in the UV limit.

The integrals $\mathcal{I}_{\alpha \beta}^{+-}(z)$ admit a closed-form expression only for $\alpha=\beta=1$. Indeed, by performing a double integration by parts:
\begin{align}
\mathcal{I}_{1,1}^{+-}(z) &= \int_0^\infty \mathrm{d}\vartheta \frac{\sinh \vartheta \cosh \vartheta}{1 + e^{z\cosh \vartheta}} = \sum_{n=1}^{\infty} (-1)^{n+1} \int_0^\infty \mathrm{d}\vartheta \, \cosh \vartheta \sinh \vartheta e^{-n z \cosh \vartheta} \nonumber \\
&= \sum_{n=1}^{\infty} (-1)^{n+1} \l-\frac{1}{nz} \r \int_0^\infty \mathrm{d}\vartheta \, \cosh \vartheta \frac{\partial}{\partial \vartheta}( e^{-n z \cosh \vartheta}) \nonumber \\
&= \sum_{n=1}^{\infty} (-1)^{n+1} \l-\frac{1}{nz} \r \left[ - e^{-nz} - \int_0^\infty \mathrm{d}\vartheta \, \sinh \vartheta  e^{-n z \cosh \vartheta}\right] \nonumber \\
&= \sum_{n=1}^{\infty} (-1)^{n+1} \l-\frac{1}{nz} \r \left[ - e^{-nz} +\frac{1}{nz} \int_0^\infty \mathrm{d}\vartheta \, \frac{\partial}{\partial \vartheta}( e^{-n z\cosh \vartheta})\right] \nonumber \\
&= \sum_{n=1}^{\infty} (-1)^{n+1} \frac{1}{nz} \l 1 + \frac{1}{nz}\r e^{-nz} \nonumber \\
&= \frac{1}{z}\sum_{n=1}^{\infty} (-1)^{n+1} \frac{(e^{-z})^n}{n}  -  \frac{1}{z^2}\sum_{n=1}^{\infty}  \frac{(-e^{-z})^n}{n^2} \nonumber \\ &=
\frac{\ln(1+e^{-z})}{z} - \frac{\text{Li}_2(-e^{-z})}{z^2}.
\end{align}
This gives the free fermion average energy current \eqref{ff_energy_current}, and the small-$z$ expansion:
\be
\frac{\ln(1+e^{-z})}{z} - \frac{\text{Li}_2(-e^{-z})}{z^2} = \frac{\pi^2}{12 z^2} - \frac{1}{4} + \mathcal{O}(z),
\ee
reproduces the asymptotics \eqref{ff_energy_current_asymptotics}.
The small-$z$ asymptotics of the other $\mathcal{I}_{\alpha \beta}^{+-}(z)$ integrals can be worked out from elementary manipulations of hyperbolic functions. Let us focus for instance on the quantity  $\mathcal{I}_{1,s}^{+-}(z)$. By performing the usual expansion \eqref{geometric series} of the geometric series, we can limit ourselves to the integral:
\begin{align}
\mathcal{J}(s,z) &\equiv \int_{0}^\infty \mathrm{d}\vartheta\, \cosh \vartheta \sinh (s \vartheta) e^{-z \cosh \vartheta} \nonumber \\&= \int_{0}^\infty \mathrm{d}\vartheta\, \cosh \vartheta \cosh (s \vartheta) e^{-z \cosh \vartheta} - \int_{0}^\infty \mathrm{d}\vartheta\, \cosh \vartheta e^{-s\vartheta -z \cosh \vartheta},
\end{align}
for $z > 0$. Since
\be
0 < \int_{0}^\infty \mathrm{d}\vartheta\, \cosh \vartheta e^{-s\vartheta -z \cosh \vartheta} < K_1 (z),
\ee
the following bounds hold:
\be
K_{s+1}(z)-\frac{s}{z}K_s(z) - K_1(z) \,<\, \mathcal{J}(s, z) \,<\, K_{s+1}(z)-\frac{s}{z}K_s(z),
\ee
and therefore
\be
\frac{2^{s-1}\Gamma(s+1)}{z^{s+1}}  - \frac{1}{z}\,\lesssim \,\mathcal{J}(s, z) \,\lesssim\, \frac{2^{s-1}\Gamma(s+1)}{z^{s+1}}, \quad \text{as} \,\, z \to 0^+.
\ee
This implies that $\mathcal{I}_{1,s}^{+-}(z)$ and $\mathcal{I}_{1,s}^{++}(z)$ have the same small-$z$ asymptotics, and similar bounds can be found for all the other functions $\mathcal{I}_{\alpha \beta}^{+-}(z)$.
\printbibliography

@article{Zamolodchikov:1989hfa,
    author = "Zamolodchikov, A. B.",
    editor = "Jimbo, M. and Miwa, T. and Tsuchiya, A.",
    title = "{Integrable field theory from conformal field theory}",
    journal = "Adv. Stud. Pure Math.",
    volume = "19",
    pages = "641--674",
    year = "1989"
}

@article{staircase,
    author = "Mazzoni, Michele and Pomponio, Octavio and Castro-Alvaredo, Olalla A. and Ravanini, Francesco",
    title = "{The staircase model: massless flows and hydrodynamics}",
    eprint = "2105.13349",
    archivePrefix = "arXiv",
    primaryClass = "hep-th",
    doi = "10.1088/1751-8121/ac2141",
    journal = "J. Phys. A",
    volume = "54",
    number = "40",
    pages = "404005",
    year = "2021"
}

@Article{Zamolodchikov1991tba,
     author    = "Zamolodchikov, Al.B.",
     title     = "{Thermodynamic Bethe ansatz in relativistic models. Scaling three state Potts
     and Lee-Yang models}",
    doi = "10.1016/0550-3213(90)90333-9",
     journal   = "Nucl. Phys.",
     volume    = "B342",
     year      = "1990",
     pages     = "695-720",
     SLACcitation  = "%%CITATION = NUPHA,B342,695;%%"
}

@Article{Klassen1991,
     author    = "Klassen, T. R. and Melzer, E.",
     title     = "{The Thermodynamics of purely elastic scattering theories
                  and conformal perturbation theory}",
     doi = "10.1016/0550-3213(90)90643-R",
     journal   = "Nucl. Phys.",
     volume    = "B350",
     year      = "1991",
     pages     = "635-689",
     SLACcitation  = "%%CITATION = NUPHA,B350,635;%%"
}

@article{klassen1990,
  title={Purely elastic scattering theories and their ultraviolet limits},
  author={Klassen, Timothy R and Melzer, Ezer},
  journal={Nuclear Physics B},
  volume={338},
  number={3},
  pages={485--528},
  year={1990},
  publisher={Elsevier}
}

@article{GGE_definition,
  title = {Relaxation in a Completely Integrable Many-Body Quantum System: An Ab Initio Study of the Dynamics of the Highly Excited States of 1D Lattice Hard-Core Bosons},
  author = {Rigol, Marcos and Dunjko, Vanja and Yurovsky, Vladimir and Olshanii, Maxim},
  journal = {Phys. Rev. Lett.},
  volume = {98},
  issue = {5},
  pages = {050405},
  numpages = {4},
  year = {2007},
  month = {Feb},
  publisher = {American Physical Society},
  doi = {10.1103/PhysRevLett.98.050405},
  url = {https://link.aps.org/doi/10.1103/PhysRevLett.98.050405}
}

@article{Zamolodchikov:2004ce,
    author = "Zamolodchikov, Alexander B.",
    title = "{Expectation value of composite field T anti-T in two-dimensional quantum field theory}",
    eprint = "hep-th/0401146",
    archivePrefix = "arXiv",
    reportNumber = "BONN-TH-2004-02",
    month = "1",
    year = "2004"
}

@article{Cavaglia:2016oda,
    author = "Cavagli\`a, Andrea and Negro, Stefano and Sz\'ecs\'enyi, Istv\'an M. and Tateo, Roberto",
    title = "{$T \bar{T}$-deformed 2D Quantum Field Theories}",
    eprint = "1608.05534",
    archivePrefix = "arXiv",
    primaryClass = "hep-th",
    doi = "10.1007/JHEP10(2016)112",
    journal = "JHEP",
    volume = "10",
    pages = "112",
    year = "2016"
}

@article{Medenjak:2020ppv,
    author = "Medenjak, Marko and Policastro, Giuseppe and Yoshimura, Takato",
    title = "{$T\bar{T}$-Deformed Conformal Field Theories out of Equilibrium}",
    eprint = "2011.05827",
    archivePrefix = "arXiv",
    primaryClass = "cond-mat.stat-mech",
    doi = "10.1103/PhysRevLett.126.121601",
    journal = "Phys. Rev. Lett.",
    volume = "126",
    number = "12",
    pages = "121601",
    year = "2021"
}

@article{Medenjak:2020bpe,
    author = "Medenjak, Marko and Policastro, Giuseppe and Yoshimura, Takato",
    title = "{Thermal transport in $T \bar{T}$-deformed conformal field theories: From integrability to holography}",
    eprint = "2010.15813",
    archivePrefix = "arXiv",
    primaryClass = "cond-mat.stat-mech",
    doi = "10.1103/PhysRevD.103.066012",
    journal = "Phys. Rev. D",
    volume = "103",
    number = "6",
    pages = "066012",
    year = "2021"
}

@article{Smirnov:2016lqw,
    author = "Smirnov, F. A. and Zamolodchikov, A. B.",
    title = "{On space of integrable quantum field theories}",
    eprint = "1608.05499",
    archivePrefix = "arXiv",
    primaryClass = "hep-th",
    doi = "10.1016/j.nuclphysb.2016.12.014",
    journal = "Nucl. Phys. B",
    volume = "915",
    pages = "363--383",
    year = "2017"
}

@article{ESSLER_ghd_review,
title = {A short introduction to Generalized Hydrodynamics},
journal = {Physica A: Statistical Mechanics and its Applications},
pages = {127572},
year = {2022},
issn = {0378-4371},
doi = {https://doi.org/10.1016/j.physa.2022.127572},
url = {https://www.sciencedirect.com/science/article/pii/S0378437122003971},
author = {Fabian H.L. Essler}
}

@article{castro2016emergent,
    author = "Castro-Alvaredo, Olalla A. and Doyon, Benjamin and Yoshimura, Takato",
    title = "{Emergent hydrodynamics in integrable quantum systems out of equilibrium}",
    eprint = "1605.07331",
    archivePrefix = "arXiv",
    primaryClass = "cond-mat.stat-mech",
    doi = "10.1103/PhysRevX.6.041065",
    journal = "Phys. Rev. X",
    volume = "6",
    number = "4",
    pages = "041065",
    year = "2016"
}

@article{bernard2012energy,
  title={Energy flow in non-equilibrium conformal field theory},
  author={Bernard, Denis and Doyon, Benjamin},
  journal={J. Phys. A},
doi = "10.1088/1751-8113/45/36/362001",
  volume={45},
  number={36},
  pages={362001},
  year={2012},
  publisher={IOP Publishing}
}

@article{affleck1988universal,
    author = "Affleck, Ian",
    title = "{Universal Term in the Free Energy at a Critical Point and the Conformal Anomaly}",
    reportNumber = "Print-86-0085 (PRINCETON)",
    doi = "10.1103/PhysRevLett.56.746",
    journal = "Phys. Rev. Lett.",
    volume = "56",
    pages = "746--748",
    year = "1986"
}

@article{blote1986conformal,
    author = "Bloete, H. W. J. and Cardy, John L. and Nightingale, M. P.",
    title = "{Conformal Invariance, the Central Charge, and Universal Finite Size Amplitudes at Criticality}",
    doi = "10.1103/PhysRevLett.56.742",
    journal = "Phys. Rev. Lett.",
    volume = "56",
    pages = "742--745",
    year = "1986"
}

@article{DynkinTBA,
author = {Ravanini, F. and Valleriani, A. and Tateo, R.},
title = "{Dynkin TBA’s}",
journal = {Int. J. Mod. Phys. A},
volume = {08},
number = {10},
pages = {1707-1727},
year = {1993},
doi = {10.1142/S0217751X93000709},
URL = { 
        https://doi.org/10.1142/S0217751X93000709
},
}

@article{bazhanov1996integrable,
    author = "Bazhanov, Vladimir V. and Lukyanov, Sergei L. and Zamolodchikov, Alexander B.",
    title = "{Integrable structure of conformal field theory, quantum KdV theory and thermodynamic Bethe ansatz}",
    eprint = "hep-th/9412229",
    archivePrefix = "arXiv",
    reportNumber = "CLNS-94-1316, RU-94-98",
    doi = "10.1007/BF02101898",
    journal = "Commun. Math. Phys.",
    volume = "177",
    pages = "381--398",
    year = "1996"
}

@article{zamolodchikov2006resonance,
  title={Resonance factorized scattering and roaming trajectories},
  author={Zamolodchikov, Al B},
  journal={J. Phys. A},
    doi = "10.1088/0305-4470/39/41/S08",
  volume={39},
  number={41},
  pages={12847},
  year={2006},
  publisher={IOP Publishing}
}

@article{bernard2016conformal,
  title={Conformal field theory out of equilibrium: a review},
  author={Bernard, Denis and Doyon, Benjamin},
  journal={J. Stat. Mech.},
  doi = "10.1088/1742-5468/2016/06/064005",
  volume={2016},
  number={6},
  pages={064005},
  year={2016},
  publisher={IOP Publishing}
}

@article{doyon2017note,
    author = "Doyon, Benjamin and Yoshimura, Takato",
    title = "{A note on generalized hydrodynamics: inhomogeneous fields and other concepts}",
    eprint = "1611.08225",
    archivePrefix = "arXiv",
    primaryClass = "cond-mat.stat-mech",
    doi = "10.21468/SciPostPhys.2.2.014",
    journal = "SciPost Phys.",
    volume = "2",
    number = "2",
    pages = "014",
    year = "2017"
}

@article{bertini_ghd,
  title = {Transport in Out-of-Equilibrium $XXZ$ Chains: Exact Profiles of Charges and Currents},
  author = {Bertini, Bruno and Collura, Mario and De Nardis, Jacopo and Fagotti, Maurizio},
  journal = {Phys. Rev. Lett.},
  volume = {117},
  issue = {20},
  pages = {207201},
  numpages = {8},
  year = {2016},
  month = {Nov},
  publisher = {American Physical Society},
  doi = {10.1103/PhysRevLett.117.207201},
  url = {https://link.aps.org/doi/10.1103/PhysRevLett.117.207201}
}

@article{yoshimura2020collision,
    author = "Yoshimura, Takato and Spohn, Herbert",
    title = "{Collision rate ansatz for quantum integrable systems}",
    eprint = "2004.07113",
    archivePrefix = "arXiv",
    primaryClass = "cond-mat.stat-mech",
    doi = "10.21468/SciPostPhys.9.3.040",
    journal = "SciPost Phys.",
    volume = "9",
    number = "3",
    pages = "040",
    year = "2020"
}

@article{bajnok2020exact,
    author = "Bajnok, Zoltan and Vona, Istvan",
    title = "{Exact finite volume expectation values of conserved currents}",
    eprint = "1911.08525",
    archivePrefix = "arXiv",
    primaryClass = "hep-th",
    doi = "10.1016/j.physletb.2020.135446",
    journal = "Phys. Lett. B",
    volume = "805",
    pages = "135446",
    year = "2020"
}

@article{borsi2021current,
    author = "Borsi, M. \'arton and Pozsgay, Bal\'azs and Pristy\'ak, Levente",
    title = "{Current operators in integrable models: a review}",
    eprint = "2103.12160",
    archivePrefix = "arXiv",
    primaryClass = "cond-mat.stat-mech",
    doi = "10.1088/1742-5468/ac0f6b",
    journal = "J. Stat. Mech.",
    volume = "2109",
    pages = "094001",
    year = "2021"
}

@article{borsi2020current,
  title={Current operators in Bethe ansatz and generalized hydrodynamics: An exact quantum-classical correspondence},
  author={Borsi, M{\'a}rton and Pozsgay, Bal{\'a}zs and Pristy{\'a}k, Levente},
url={http://dx.doi.org/10.1103/PhysRevX.10.011054},
   DOI={10.1103/physrevx.10.011054},
  journal={Phys. Rev. X},
  volume={10},
  number={1},
  pages={011054},
  year={2020},
  publisher={APS}
}

@article{kinoshita2006quantum,
  title={A quantum Newton's cradle},
  author={Kinoshita, Toshiya and Wenger, Trevor and Weiss, David S},
  journal={Nature},
  volume={440},
  number={7086},
  DOI = {https://doi.org/10.1038/nature04693},
  pages={900--903},
  year={2006},
  publisher={Nature Publishing Group UK London}
}

@article{failure,
   title={Correlations after Quantum Quenches in the XXZ Spin Chain: Failure of the Generalized Gibbs Ensemble},
   volume={113},
   ISSN={1079-7114},
   url={http://dx.doi.org/10.1103/PhysRevLett.113.117203},
   DOI={10.1103/physrevlett.113.117203},
   number={11},
   journal={Phys. Rev. Lett.},
   publisher={American Physical Society (APS)},
   author={Pozsgay, B. and Mestyán, M. and Werner, M. A. and Kormos, M. and Zaránd, G. and Takács, G.},
   year={2014},
   month=sep }

@article{Fioretto:2009yq,
    author = "Fioretto, Davide and Mussardo, Giuseppe",
    title = "{Quantum Quenches in Integrable Field Theories}",
    eprint = "0911.3345",
    archivePrefix = "arXiv",
    primaryClass = "cond-mat.stat-mech",
    doi = "10.1088/1367-2630/12/5/055015",
    journal = "New J. Phys.",
    volume = "12",
    pages = "055015",
    year = "2010"
}

@article{Zamolodchikov:1991vx,
    author = "Zamolodchikov, A. B.",
    title = "{From tricritical Ising to critical Ising by thermodynamic Bethe ansatz}",
    reportNumber = "ENS-LPS-327-1991",
    doi = "10.1016/0550-3213(91)90423-U",
    journal = "Nucl. Phys. B",
    volume = "358",
    pages = "524--546",
    year = "1991"
}

@article{Lorenzo2,
    author = "Bertini, Bruno and Piroli, Lorenzo",
    title = "{Low-temperature transport in out-of-equilibrium XXZ chains}",
    eprint = "1711.00519",
    archivePrefix = "arXiv",
    primaryClass = "cond-mat.stat-mech",
    doi = "10.1088/1742-5468/aab04b",
    journal = "J. Stat. Mech.",
    volume = "1803",
    number = "3",
    pages = "033104",
    year = "2018"
}

@article{LucaBe,
  title = {Nonequilibrium thermal transport in the quantum Ising chain},
  author = {De Luca, Andrea and Viti, Jacopo and Bernard, Denis and Doyon, Benjamin},
  journal = {Phys. Rev. B},
  volume = {88},
  issue = {13},
  pages = {134301},
  numpages = {6},
  year = {2013},
  month = {Oct},
  publisher = {American Physical Society},
  doi = {10.1103/PhysRevB.88.134301},
  url = {https://link.aps.org/doi/10.1103/PhysRevB.88.134301}
}

@ARTICLE{NatureBe,
       author = {{Bhaseen}, M.~J. and {Doyon}, Benjamin and {Lucas}, Andrew and {Schalm}, Koenraad},
        title = "{Energy flow in quantum critical systems far from equilibrium}",
      journal = {Nature Physics},
         year = 2015,
        month = jun,
       volume = {11},
       number = {6},
        pages = {509-514},
          doi = {10.1038/nphys3320},
       adsurl = {https://ui.adsabs.harvard.edu/abs/2015NatPh..11..509B}
}

@article{Perfect,
  title = {Ballistic front dynamics after joining two semi-infinite quantum Ising chains},
  author = {Perfetto, Gabriele and Gambassi, Andrea},
  journal = {Phys. Rev. E},
  volume = {96},
  issue = {1},
  pages = {012138},
  numpages = {22},
  year = {2017},
  month = {Jul},
  publisher = {American Physical Society},
  doi = {10.1103/PhysRevE.96.012138},
  url = {https://link.aps.org/doi/10.1103/PhysRevE.96.012138}
}

@Article{Marton,
	title={{Inhomogeneous quenches in the transverse field Ising chain: scaling and front dynamics}},
	author={Márton Kormos},
	journal={SciPost Phys.},
	volume={3},
	pages={020},
	year={2017},
	publisher={SciPost},
	doi={10.21468/SciPostPhys.3.3.020},
	url={https://scipost.org/10.21468/SciPostPhys.3.3.020},
}

@article{Luca2,
  title = {Energy transport between two integrable spin chains},
  author = {Biella, Alberto and De Luca, Andrea and Viti, Jacopo and Rossini, Davide and Mazza, Leonardo and Fazio, Rosario},
  journal = {Phys. Rev. B},
  volume = {93},
  issue = {20},
  pages = {205121},
  numpages = {13},
  year = {2016},
  month = {May},
  publisher = {American Physical Society},
  doi = {10.1103/PhysRevB.93.205121},
  url = {https://link.aps.org/doi/10.1103/PhysRevB.93.205121}
}

@article{Luca,
  title = {Energy transport in Heisenberg chains beyond the Luttinger liquid paradigm},
  author = {De Luca, Andrea and Viti, Jacopo and Mazza, Leonardo and Rossini, Davide},
  journal = {Phys. Rev. B},
  volume = {90},
  issue = {16},
  pages = {161101},
  numpages = {5},
  year = {2014},
  month = {Oct},
  publisher = {American Physical Society},
  doi = {10.1103/PhysRevB.90.161101},
  url = {https://link.aps.org/doi/10.1103/PhysRevB.90.161101}
}

@article{KM,
  title = {Nonequilibrium thermal transport and its relation to linear response},
  author = {Karrasch, C. and Ilan, R. and Moore, J. E.},
  journal = {Phys. Rev. B},
  volume = {88},
  issue = {19},
  pages = {195129},
  numpages = {9},
  year = {2013},
  month = {Nov},
  publisher = {American Physical Society},
  doi = {10.1103/PhysRevB.88.195129},
  url = {https://link.aps.org/doi/10.1103/PhysRevB.88.195129}
}

@article{Bernard_2014,
   title={Non-Equilibrium Steady States in Conformal Field Theory},
   volume={16},
   ISSN={1424-0661},
   url={http://dx.doi.org/10.1007/s00023-014-0314-8},
   DOI={10.1007/s00023-014-0314-8},
   number={1},
   journal={Annales Henri Poincaré},
   publisher={Springer Science and Business Media LLC},
   author={Bernard, Denis and Doyon, Benjamin},
   year={2014},
   month=jan, pages={113–161} }

@article{Lorenzo1,
    author = "Bertini, Bruno and Piroli, Lorenzo and Calabrese, Pasquale",
    title = "{Universal Broadening of the Light Cone in Low-Temperature Transport}",
    eprint = "1709.10096",
    archivePrefix = "arXiv",
    primaryClass = "cond-mat.stat-mech",
    doi = "10.1103/PhysRevLett.120.176801",
    journal = "Phys. Rev. Lett.",
    volume = "120",
    number = "17",
    pages = "176801",
    year = "2018"
}

@article{ProsenXXZ,
  title = {Quasilocal Conserved Operators in the Isotropic Heisenberg Spin-$1/2$ Chain},
  author = {Ilievski, Enej and Medenjak, Marko and Prosen, T},
  journal = {Phys. Rev. Lett.},
  volume = {115},
  issue = {12},
  pages = {120601},
  numpages = {5},
  year = {2015},
  month = {Sep},
  publisher = {American Physical Society},
  doi = {10.1103/PhysRevLett.115.120601},
  url = {https://link.aps.org/doi/10.1103/PhysRevLett.115.120601}
}

@article{ilievski2016quasilocal,
  title={Quasilocal charges in integrable lattice systems},
  author={Ilievski, Enej and Medenjak, Marko and Prosen, Toma{\v{z}} and Zadnik, Lenart},
  journal={J. Stat. Mech.},
  volume={2016},
  number={6},
 doi = {10.1088/1742-5468/2016/06/064008},
  pages={064008},
  year={2016},
  publisher={IOP Publishing}
}

@article{vernier2017quasilocal,
  title={Quasilocal charges and progress towards the complete GGE for field theories with nondiagonal scattering},
  author={Vernier, Eric and Cubero, Axel Cort{\'e}s},
  journal={J. Stat. Mech.},
  volume={2017},
   doi = "10.1088/1742-5468/aa5288",
  number={2},
  pages={023101},
  year={2017},
  publisher={IOP Publishing}
}

@article{Mossel2012generalized,
    author = "Mossel, Jorn and Caux, Jean-Sebastien",
    title = "{Generalized TBA and generalized Gibbs}",
    eprint = "1203.1305",
    archivePrefix = "arXiv",
    primaryClass = "cond-mat.quant-gas",
    doi = "10.1088/1751-8113/45/25/255001",
    journal = "J. Phys. A",
    volume = "45",
    pages = "255001",
    year = "2012"
}

@article{doyon2020lecture,
    author = "Doyon, Benjamin",
    title = "{Lecture notes on Generalised Hydrodynamics}",
    eprint = "1912.08496",
    archivePrefix = "arXiv",
    primaryClass = "cond-mat.stat-mech",
    doi = "10.21468/SciPostPhysLectNotes.18",
    journal = "SciPost Phys. Lect. Notes",
    volume = "18",
    pages = "1",
    year = "2020"
}

@article{cardy2010ubiquitous,
   title={The ubiquitous ‘c’: from the Stefan–Boltzmann law to quantum information},
   volume={2010},
   ISSN={1742-5468},
   url={http://dx.doi.org/10.1088/1742-5468/2010/10/P10004},
   DOI={10.1088/1742-5468/2010/10/p10004},
   number={10},
   journal={J. Stat. Mech.},
   publisher={IOP Publishing},
   author={Cardy, John},
   year={2010},
   month=oct, pages={P10004} }

@article{zamolodchikov2006thermodynamic,
    author = "Zamolodchikov, Alexei B.",
    title = "{On the thermodynamic Bethe ansatz equation in sinh-Gordon model}",
    eprint = "hep-th/0005181",
    archivePrefix = "arXiv",
    reportNumber = "LPM-00-15",
    doi = "10.1088/0305-4470/39/41/S09",
    journal = "J. Phys. A",
    volume = "39",
    pages = "12863--12887",
    year = "2006"
}

@article{cardy1989s,
    author = "Cardy, John L. and Mussardo, G.",
    title = "{S Matrix of the Yang-Lee Edge Singularity in Two-Dimensions}",
    reportNumber = "UCSB-TH-89-11",
    doi = "10.1016/0370-2693(89)90818-6",
    journal = "Phys. Lett. B",
    volume = "225",
    pages = "275--278",
    year = "1989"
}

@article{OurUnpublished,
  author={Travaglino, Riccardo and Mazzoni, Michele and Castro-Alvaredo, Olalla A.},
  title ={Generalised Hydrodynamics of $\mathrm{T \bar{T}}$-Deformed Integrable Quantum Field Theories},
  journal={To appear soon},
}

@article{zamolodchikov1991ADE,
  title={On the thermodynamic Bethe ansatz equations for reflectionless ADE scattering theories},
  author={Zamolodchikov, Al B},
  journal={Physics Letters B},
  volume={253},
  number={3-4},
  pages={391--394},
  year={1991},
  publisher={North-Holland}
}

@article{myers2020transport,
   title={Transport fluctuations in integrable models out of equilibrium},
   volume={8},
   ISSN={2542-4653},
   url={http://dx.doi.org/10.21468/SciPostPhys.8.1.007},
   DOI={10.21468/scipostphys.8.1.007},
   number={1},
   journal={SciPost Physics},
   publisher={Stichting SciPost},
   author={Myers, Jason and Bhaseen, Joe and Harris, Rosemary J. and Doyon, Benjamin},
   year={2020},
   month=jan }

@article{horvath2019hydrodynamics,
  title={Hydrodynamics of massless integrable RG flows and a non-equilibrium c-theorem},
  author={Horv{\'a}th, D{\'a}vid X},
  journal={Journal of High Energy Physics},
  volume={2019},
  number={10},
  pages={1--40},
  year={2019},
  publisher={Springer}
}

@article{castro2014thermodynamic,
  title={Thermodynamic Bethe ansatz for non-equilibrium steady states: exact energy current and fluctuations in integrable QFT},
  author={Castro-Alvaredo, Olalla and Chen, Yixiong and Doyon, Benjamin and Hoogeveen, Marianne},
  journal={Journal of Statistical Mechanics: Theory and Experiment},
  volume={2014},
  number={3},
  pages={P03011},
  year={2014},
  publisher={IOP Publishing}
}
\end{document}